\documentclass[aps, prb, reprint, superscriptaddress]{revtex4-2}

\usepackage[letterpaper,left=1in,right=1in,top=1in,bottom=1in]{geometry}

\usepackage{amsmath,amssymb,bm}
\usepackage{graphicx}
\usepackage[T1]{fontenc}
\usepackage{fourier}
\usepackage{baskervald}

\usepackage{setspace}

\DeclareMathOperator{\re}{Re}
\DeclareMathOperator{\im}{Im}
\newcommand{\unit}[1]{\,\mathrm{#1}} 

\renewcommand{\vec}{\mathbf}

\usepackage[dvipsnames]{xcolor}
\definecolor{mycolor}{RGB}{0,0,0}

\color{mycolor}

\usepackage{hyperref}
\hypersetup{%
	pdftitle={Modeling of Plasmonic and Polaritonic Effects in Photocurrent Nanoscopy}, %
	pdfauthor={Andrey Rikhter, D. N. Basov, ..., M. M. Fogler},%
	pdfpagemode={UseNone},%
	pdfstartview={FitH},%
	breaklinks=true,%
	citecolor=blue,%
	colorlinks=true,%
	linkcolor=blue,%
	urlcolor=blue}

\date{\today}

\begin{document}
\begin{abstract}

We present a basic framework for modeling collective mode effects
in photocurrent measurements performed on two-dimensional materials using nano-optical scanned probes.
We consider photothermal, photovoltaic, and bolometric contributions to the photocurrent.
We show that any one of these can dominate depending on frequency, temperature, applied bias, and sample geometry. 
Our model is able to account for periodic spatial oscillations (fringes) of the photocurrent observed near sample edges or inhomogeneities. 
For the case of a non-absorbing substrate,
we find a direct relation between the
spectra measured by the photocurrent nanoscopy and its parental scanning technique,
near-field optical microscopy.

\end{abstract}

\title{Modeling of Plasmonic and Polaritonic Effects in Photocurrent Nanoscopy}

\author{A. Rikhter}
\affiliation{Department of Physics, University of California San Diego, 9500 Gilman Drive, La Jolla, California 92093}

\author{D. N. Basov}
\affiliation{Department of Physics, Columbia University, New York, New York 10027}

\author{M. M. Fogler}
\affiliation{Department of Physics, University of California San Diego, 9500 Gilman Drive, La Jolla, California 92093}

\maketitle

\section{Introduction}
\label{sec:intro} 

\begin{figure}[thb]
	\includegraphics[width=2.50 in]{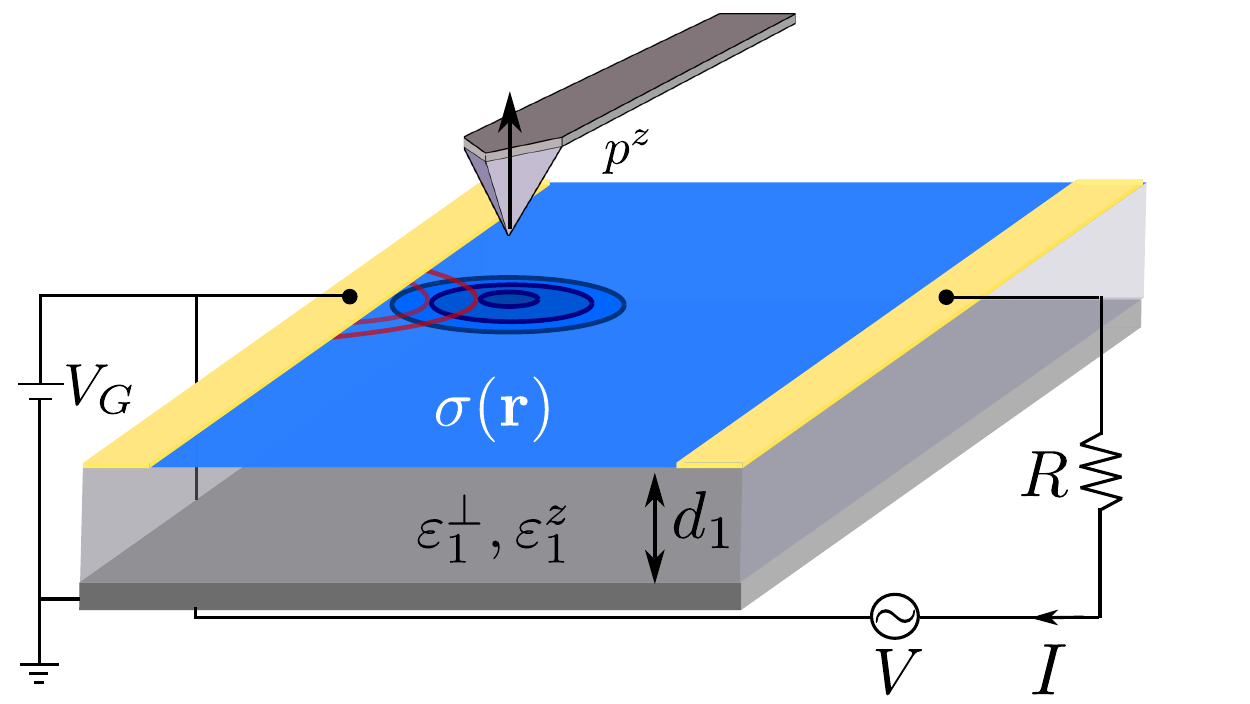}
	\caption{Sketch of a theoretical model for photocurrent nanoscopy.
	The tip of a scanned probe brought near the sample possesses a dipole moment of amplitude $p^z$ induced by a focused light beam. The sample consists of a graphene sheet of conductivity $\sigma(\vec{r})$ placed on a substrate made of dielectric layers with in/out-of-plane permittivities $\varepsilon^\perp_i(\omega)$, $\varepsilon^z_i(\omega)$.
	The locally enhanced electric field modifies the current $I$ through the resistor $R$ due to the presence of the probe tip. 
	}
	\label{fig:heterostructure_schematic} 
\end{figure}

Scanning photocurrent microscopy is traditionally performed using a
focused light beam~\cite{Kallmann1960, Guettler1970, Marek1984, GRAHAM2013}.
In a modern variant of this technique,
the focusing of incident light is achieved instead by a sharp metal tip,
as illustrated schematically in Fig.~\ref{fig:heterostructure_schematic}.
Such a tip acts as an optical antenna that
couples a locally enhanced near-field to free-space radiation.
In experiment, the tip is scanned and the dc photocurrent current generated in the sample
is measured as a function of the tip position
using electric contacts positioned somewhere on the sample periphery.
Below we refer to this technique as scanning near-field photocurrent microscopy
or photocurrent nanoscopy.
The instrumentation involved in such measurements
can also be utilized to perform scattering-type scanning near-field optical microscopy (s-SNOM).
In s-SNOM one detects light scattered by the tip instead of the photocurrent.
In practice, s-SNOM and photocurrent nanoscopy are performed together,
providing complementary information about the system.
This combination of techniques has been successfully applied to
probe graphene and other two-dimensional (2D) materials~\cite{Basov2014, Basov2016, Ni2018, Lundeberg2017, Hesp2021}
demonstrating spatial resolution of $\sim 20\,\mathrm{nm}$,
which is
orders of magnitude better than the diffraction-limited traditional approach.

Recent photocurrent nanoscopy experiments revealed
distinctive spectral resonances and periodic interference patterns occurring near sample edges and inhomogeneities~\cite{Basov2016, Ni2018, Lundeberg2017, Hesp2021}.
These features have been attributed to collective modes, plasmon- and phonon-polaritons,
excited in graphene and underlying 2D substrate materials.
In this paper, we aim to formulate a theoretical model for such collective mode effects.

Whereas modeling of s-SNOM has been actively pursued in the past decade~\cite{Cvitkovic2007, McLeod2014, Jiang2016, Chen2021}, photocurrent nanoscopy has received less attention. 
Theoretical analysis of the latter is more effortful because
in addition to the electromagnetic tip-sample coupling,
one also has to account for multiple possible mechanisms of the DC photocurrent generation.
We focus on the case where the photocurrent scales linearly with the incident light intensity, i.e.,
as a second power of the in-plane AC electric field $\vec{E}(\vec{r}) e^{-i \omega t} + \mathrm{c.c}$.
Assuming the system contains only inversion-symmetric materials,
such a second-order nonlinear effect can arise if the inversion symmetry
is violated by boundary conditions, structural defects, or externally applied fields.
For example,
nonvanishing photocurrent can exist if the 
carrier density $n(\vec{r})$ in the
scanned region is nonuniform.
Photocurrent can also be generated if the magnitude or the phase 
of $\vec{E}(\vec{r})$ is spatially dependent
(the latter corresponds to a nonzero in-plane momentum)~\cite{Glazov2014}.
If there exists a DC electric field $\vec{E}^\mathrm{DC}(\vec{r})$ in the system,
the photocurrent can include terms that scale as $E^\mathrm{DC} |\vec{E}|^2$,
which we also consider in our calculations.
Altogether we examine three mechanisms of photocurrent generation:
bolometric (BM), photothermal (PT), and photovoltaic (PV).
The relative importance of the BM, PT, and PV effects depends on a system.
In common bulk semiconductors, the Joule heating of charge carriers is suppressed by an efficient cooling by optical phonons~\cite{Seeger2004}. 
The resultant PT current is small and the photocurrent is mostly due to the PV effect. 
In graphene, the linear quasiparticle dispersion combined with the high optical phonon frequency inhibits electron cooling, which enhances the PT contribution~\cite{Bistritzer2009, Song2011}.
Experiments performed on graphene $p$-$n$ junctions have demonstrated that in the absence of a bias, $V = 0$, the PT effect is typically the dominant photocurrent mechanism~\cite{Gabor2011}.
However, the BM current quickly becomes
the largest contribution as the bias $V$ is increased from zero~\cite{Freitag_2012}.

As we show in the remainder of this paper, collective mode effects in photocurrent, which is the main target of our investigation,
can manifest themselves via all these three mechanisms.
In Sec.~\ref{sec:mechanisms},
we introduce basic definitions and relations concerning the BM, PT, and PV effects.
In Sec.~\ref{sec:spatial},
we present illustrative examples how collective modes --- plasmons and polaritons ---
can generate characteristic spatial patterns detectable in photocurrent nanoscopy.
In Sec.~\ref{sec:frequency} we study
the collective mode signatures in the frequency dependence of the PT photocurrent.
We present main equations of our model in Sec.~\ref{sec:equations}.
Further technical details are given in the Appendix.

\section{Definitions of photovoltages, currents, and conductivities}
\label{sec:mechanisms} 

The model system we study is shown schematically in Fig.~\ref{fig:heterostructure_schematic}.
The measured quantity is the total current $I$ given by
\begin{equation}
	I = \frac{V + V^{\mathrm{PH}}}{R_g + R}, 
	\label{eqn:I_aux}
\end{equation}
where $V$ is the bias voltage applied between the contacts, $V^{\mathrm{PH}}$ is the photovoltage (discussed below),
$R_g$ is the sample resistance in the dark,
and $R$ is any additional resistance in series with $R_g$, e.g., the contact resistance.
Assuming $R$ remains constant under illumination, the photocurrent is proportional to $V^{\mathrm{PH}}$,
which is in turn proportional to the local photoinduced electromotive force (EMF) $\vec{F}^{\mathrm{PH}}(\vec{r})$.
If $\vec{F}^{\mathrm{PH}}(\vec{r})$,
the EMF in the dark
\begin{equation}
	F_i^\mathrm{DC}(\vec{r})
	= E_i^\mathrm{DC}(\vec{r}) - \frac{1}{e}\, \partial_i \mu^\mathrm{DC}(\vec{r}),
	\quad i \in \{x, y\},
	\label{eqn:F_DC}  
\end{equation}
and the linear-response DC conductivity
$\sigma^\mathrm{DC}$ were all uniform, and the sample were
infinite,
the two EMFs would be simply additive.
(Here $\mu^\mathrm{DC}$ is the equilibrium distribution of the chemical potential and
$e = -|e|$ is the electron charge.)
The photoinduced EMF would drive an extra uniform current
\begin{equation}
	j_i^\mathrm{PH} = \sigma^\mathrm{DC} F_i^{\mathrm{PH}}\,. 
	\label{eqn:jph}  
\end{equation}
In the practice of photocurrent nanoscopy, $\vec{F}^{\mathrm{PH}}(\vec{r})$ is nonuniform,
the sample is of finite size and may have an irregular shape.
In such a case Eq.~\eqref{eqn:jph} serves as a formal definition of $j_i^\mathrm{PH}(\vec{r})$;
however, the actual current density is different because the total EMF must readjust itself to ensure current conservation in the steady state.
Nevertheless, the relation between $V^{\mathrm{PH}}$ and $\vec{F}^{\mathrm{PH}}$
(or equivalently, $\mathbf{j}^\mathrm{PH}$)
can be conveniently expressed using the Shockley-Ramo theorem (see, e.g., Ref.~\cite{Song2014}):
\begin{equation}
	V^{\mathrm{PH}} = \int d^2 r \, F_i^{\mathrm{PH}}(\vec{r}) \psi_i^\mathrm{DC}(\vec{r}),
	 \quad
	\psi_i(\vec{r}) = \frac{1}{I^\mathrm{DC}}\, j_i^\mathrm{DC}(\vec{r}). 
\label{eqn:shockley-ramo}  
\end{equation}
Here the repeated index $i$ is meant to be the summed over,
\begin{equation}
 j_i^\mathrm{DC}(\vec{r}) = \sigma^\mathrm{DC}(\vec{r})
 F_i^\mathrm{DC}(\vec{r})
 \label{eqn:j_DC}  
\end{equation}
is the current density in the dark,
and $I^\mathrm{DC}$ is the total dark current.
The auxiliary vector field $\vec{\psi}(\vec{r})$,
which has the units of inverse length in 2D,
encodes all the geometric properties of the sample and contacts and obeys
the normalization relation
\begin{equation}
	R_g = \int d^2 r \, \frac{\psi_i(\vec{r}) \psi_i(\vec{r})}{\sigma^\mathrm{DC}(\vec{r})} \,.
	\label{eqn:R_shockley-ramo}  
\end{equation}
As mentioned in Sec.~\ref{sec:intro}, we consider three contributions to the photoinduced EMF $\vec{F}^{\mathrm{PH}}$.
The first one is due to the photothermal (PT) effect:
\begin{equation}
	F_i^{\mathrm{PT}} = - S_{i j} \partial_j T, 
	\label{eqn:jPT}
\end{equation}
where $S_{i j}$ is the tensor of Seebeck coefficients (same as the thermopower tensor).
The electron temperature $T$ that enters this equation differs from the equilibrium ambient $T_0$ temperature
because of AC Joule heating.
We assume that the relation between $T - T_0$ and the heating power $\propto |E|^2$ is linear,
so that the gradient of $T$ is quadratic in the incident AC field.
Therefore, it is possible to express the PT EMF $F_i^\mathrm{PT}$ and the corresponding current density $j_i^\mathrm{PT}$ in terms of a suitable rank-three tensor $\sigma^\mathrm{PT}_{ilm}$:
\begin{align}
j_i^\mathrm{PT} \equiv \sigma^{\mathrm{DC}}	F_i^\mathrm{PT} = \sigma^\mathrm{PT}_{ilm} E_{l} E^*_{m}.
	\label{eqn:jPT_sigmaPT}
\end{align}
Since the temperature distribution is also affected by the sample geometry and
mechanisms of heat dissipation,
this tensor may depend on position $\vec{r}$ and various extrinsic factors.
The relation between $\sigma^\mathrm{PT}_{ilm}$, $S_{i j}$, and other
material properties will be further elaborated on in Sec.~\ref{sec:equations}.

The second contribution we include is due to the bolometric (BM) correction to
the DC conductivity $\sigma^\mathrm{DC}$.
The corresponding corrections to the EMF and the current density satisfy the equation
\begin{equation}
	j_i^\mathrm{BM} \equiv \sigma^{\mathrm{DC}}	
	F_i^{\mathrm{BM}} = \frac{\partial \sigma^\mathrm{DC}}{\partial T}\, (T - T_0)
	 F_i^{\mathrm{DC}}. 
	\label{eqn:jbm}
\end{equation}
The BM coefficient ${\partial \sigma^\mathrm{DC}} / {\partial T}$
is briefly discussed in Sec.~\ref{sec:spatial}. 
As with the PT photocurrent, it is in principle possible to rewrite Eq.~\eqref{eqn:jbm} in terms of a certain rank-three tensor $\sigma^\mathrm{BM}_{ilm}$ but we will not do so.

Lastly, we consider the photovoltaic (PV) current:
\begin{equation}
	j_i^\mathrm{PV} \equiv \sigma^{\mathrm{DC}}	
	F_i^{\mathrm{PV}} = \sigma_{ilm}^\mathrm{PVC} E_l E^*_m  - \frac{1}{e}\,
	\sigma^{\mathrm{DC}} \partial_i\left(\mu - \mu^{\mathrm{DC}}\right). 
	\label{eqn:jpv}
\end{equation}
(It is also referred to as the photogalvanic current in some literature.)
In Eq.~\eqref{eqn:jpv}, we split the PV response into the coherent part (the first term), due to the AC electric field, and the incoherent part (the second term), caused by the change of the chemical potential $\mu$ due to the heating $T - T_0$ and/or the photoexcited carrier density $n - n_0$.
Similar to the points made above about the PT and BM currents,
the sum of these two parts can also be written using a certain tensor $\sigma^\mathrm{PV}_{ilm}$:
\begin{align}
	j_i^\mathrm{PV} &= \sigma^\mathrm{PV}_{ilm} E_{l} E^*_{m}.
	\label{eqn:jPV_sigmaPV}
\end{align}
In turn, the full second-order conductivity tensor is the sum
\begin{equation}
	\sigma^{(2)}_{ilm} =  \sigma^\mathrm{PT}_{ilm} + \sigma^\mathrm{PV}_{ilm}.
	\label{eqn:sigma2split}
\end{equation}
Having it defined in the present work allows us make a connection with prior literature where
such a tensor has been studied~\cite{Sun2018}.
We discuss this in Sec.~\ref{sec:equations}.

\section{Photocurrent signatures of collective modes in imaging: three illustrative examples}
\label{sec:spatial}

\subsection{Hot spots due to polaritonic rays}
\label{ssec:PR}

\begin{figure}[th]
	\includegraphics[width=2.50 in]{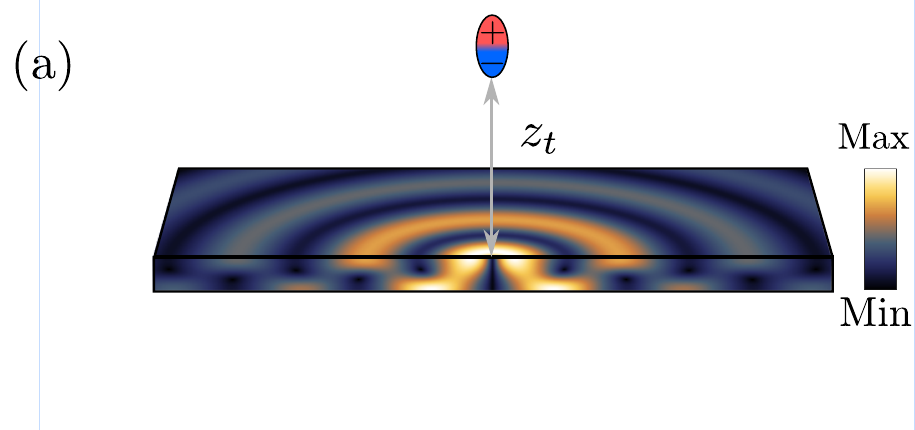}
	\includegraphics[width=2.50 in] {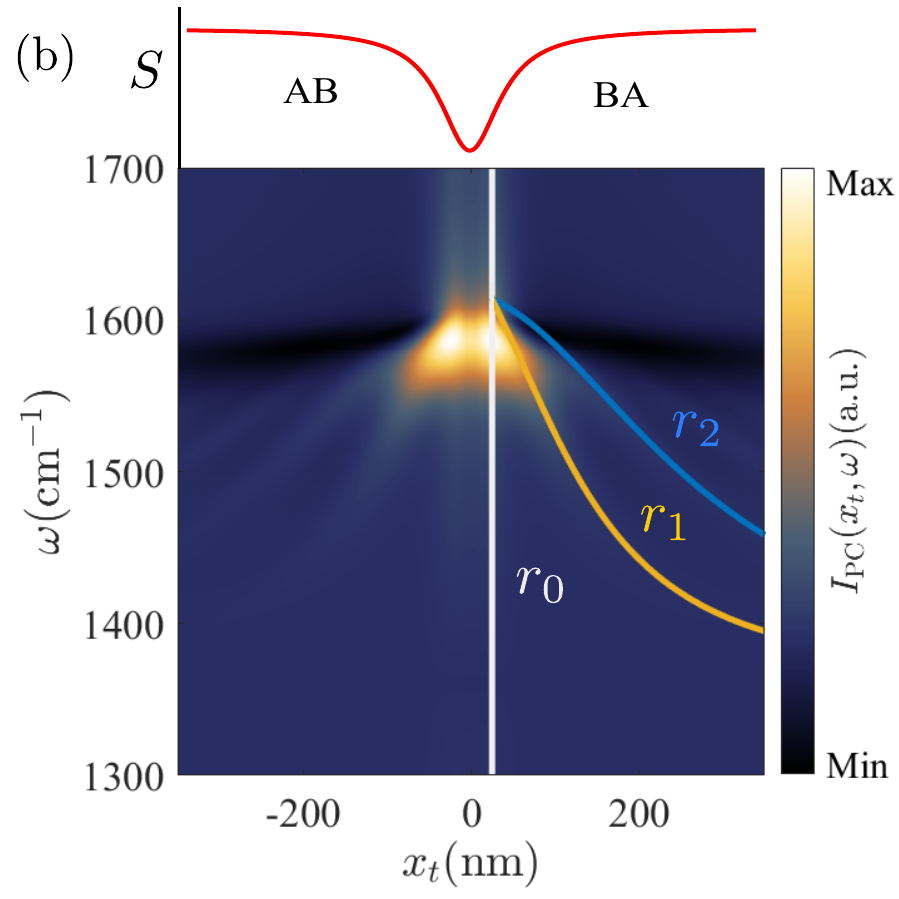}
	\caption{(a) A schematic of polariton propagation inside a slab of a hyperbolic material (HM).
		The polariton is launched by a probe, which is modeled by a point dipole a distance $z_t$ above the sample. 
		The quantity plotted is the in-plane field intensity, using the hBN optical constants from Ref.~\cite{Caldwell2014}. 
		(b) False color plot of the photocurrent as a function of the tip distance $x_t$ from the domain wall and the frequency $\omega$, showing multiple peaks inside the hyperbolic regime. 
		The tip-sample separation $z_t$ was taken to be 50 nm, equal to the hBN thickness. 
		Radii $r_k$ from Eq.~\eqref{eqn:r_n}  for $k = 0, 1, 2$ are shown with solid lines as guides to the eye.}
	\label{fig:HM_schematic}
\end{figure}

Optical phonon modes of 2D heterostructures are examples of collective excitations which may generate uncommon
effects detectable by photocurrent nanoscopy.
As one illustration, we consider 
a model system consisting of twisted bilayer graphene (TBG) deposited on a thin slab of hBN, Fig.~\ref{fig:heterostructure_schematic}.
In our previous work~\cite{Sunku2020}
the photocurrent in such a system has been experimentally found to exhibit characteristic spatial variations near the domain walls of TBG. These domain walls (also known as AB-BA boundaries or ``solitons'') form naturally in TBG if its twist angle is small enough.
Below we discuss modeling of these experiments and our theoretical predictions for observable collective mode signatures. We give some additional background necessary to understand these results but otherwise delay all the derivations until Sec.~\ref{sec:equations}.

An interesting property of hBN is its optical hyperbolicity:
the in- and out-of-plane permittivities of this material are of opposite sign in certain frequency bands,
known as Reststrahlen (RS) bands, such as $1370 < \omega < 1610\, \mathrm{cm}^{-1}$.
The optical hyperbolicity enables propagation of so-called hyperbolic phonon-polaritons inside the slab~\cite{Caldwell2014,Dai2015}. 
A localized source, such as an s-SNOM tip, typically excites several of such polariton modes simultaneously, which produces a spatial beats pattern with the period $\delta$ given by the formula
\begin{equation}
	\label{eqn:delta_hyperbolic}
	\delta = \frac{2\pi}{\Delta q} = -2 i d_1\, \frac{\sqrt{\varepsilon_1^\perp}}{\sqrt{\varepsilon_1^z}}. 
\end{equation}
Furthermore, inside the hBN slab, the electric field is strongly concentrated along certain zigzag trajectories, which we refer to as polaritonic rays, Fig.~\ref{fig:HM_schematic}(a).

The results presented below are obtained using the following additional assumptions. First, we approximate the scanned probe as a point-like dipole located at a distance $z_t$ above the sample as shown in  Fig.~\ref{fig:HM_schematic}(a). 
Such an approximation, referred to as the ``point dipole model''~\cite{Keilmann2004, Cvitkovic2007}, is commonly used in s-SNOM modeling.
Note that $z_t$ is really an adjustable parameter rather than the physical tip-sample distance. Usually it is chosen to be of the order of the curvature radius $a \sim 30\unit{nm}$ of the probe.

Second, we assume that the photocurrent is dominated by the PT effect that arises due to the local mimimum of the thermopower $S$ at the domain wall,
Fig.~\ref{fig:HM_schematic}(b). Finally, we assume that
the domain wall is infinitely long, so that the sample and therefore the photocurrent signal are translationally invariant in the longitudinal direction. 

The results obtained within this model are shown in the lower part of Fig.~\ref{fig:HM_schematic}(b).
At frequencies outside its RS band, the hBN layer is not hyperbolic and only two peaks
of photocurrent signal as a function of tip position $x_t$ are observed.
They appear when $x_t$ is approximately equal to the tip-sample separation $z_t$. 
The qualitative explanation is that at such $x_t$ the Joule heating at the domain wall is the largest,
which generates the strongest photoresponse.
Indeed, the Joule heating is proportional to the square of the in-plane electric field produced by the dipole. It vanishes directly below the dipole [which is the case in general, including the hyperbolic regime, Fig.~\ref{fig:HM_schematic}(a)] and is maximal at the lateral distance $\sim z_t$.

At frequencies inside the RS band, our calculation predicts
additional peaks of photocurrent as a function of $x_t$, which we have referred to as the ``hot spots'' in the past.
For a given $\omega$, these hot spots are separated by intervals approximately equal to $\delta$.
More precisely, they appear when the tip position $x_t$ matches the radii
\begin{equation}
	r_k = \sqrt{\frac{3}{8}(k^2 \delta^2 - z_t^2) + \frac{1}{8}\sqrt{25(k^4 \delta^4 + z_t^4) + 14 k^2 z_t^2 \delta^2}}
	\label{eqn:r_n}
\end{equation}
of ``hot rings'' created by the polaritonic rays launched by the tip [Fig.~\ref{fig:HM_schematic}(a)].
If $\delta$ is much larger than the tip-sample distance $z_t$,
our calculation predicts multiple rings to be well resolved.
At small $\delta$, the rings overlap and only weak oscillations with the period $\delta$ remain.
Only the first pair of hot spots has been observed in the experiment~\cite{Sunku2020a}. 
We expect that using cleaner samples and higher resolution probes may reveal additional ones.

\subsection{Interference fringes due to the plasmons: PT and BM effects}
\label{ssec:PSW}

Plasmons are another example of collective modes that have been imaged by both s-SNOM~\cite{Fei2012, Dai2015, Chen2012,Basov2016} and photocurrent nanoscopy~\cite{AlonsoGonzalez2016, Sunku2020a}.
Such an imaging is typically done near sample boundaries that reflect plasmons launched by the scanned probe.
Near a sample edge, the incident and reflected waves interfere, resulting in a standing-wave pattern (or ``fringes'') with spatial period ${\lambda_p} / {2}$ where $\lambda_p = {2\pi} / {q_p}$ is the plasmon wavelength and 
$q_p$ is the plasmon momentum. 
Following the structure of Sec.~\ref{ssec:PR}, below we present our modeling results for this effect but defer their derivation until Sec.~\ref{sec:equations}.
We begin the cases where the photocurrent is generated by the PT and BM mechanisms.
An example of PV plasmonic fringes is presented later in Sec.~\ref{ssec:PC}.

\subsubsection{Edge-reflected plasmons near a \texorpdfstring{$p$-$n$}{} junction}
\label{sssec:PN}

Rigorous calculation of the plasmon reflection from the edge is computationally intensive~\cite{Fei2012}.
However, there is a simple model
where the reflection from the edge ($y = 0$) is approximated by the method of images.
For the tip located at $\vec{r}_t = (x_t, y_t)$, we place an ``image'' tip at $y_i = -y_t$.
The corresponding in-plane electric field is the superposition of the source and image terms:
\begin{equation}
	\vec{E}(\vec{r}, \vec{r}_t)= \vec{E}_t(x - x_t, y - y_t) - \vec{E}_t(x - x_t, y + y_t),
	\label{eqn:E_inplane_edge}
\end{equation}
where $\vec{E}_t(x, y)$ is the in-plane field produced by the tip in an infinite uniform sample.
This approximation is reasonably accurate.
Its main deficiency concerns the position and the amplitude of the very first fringe.

\begin{figure*}
	\includegraphics[width=3.0 in]{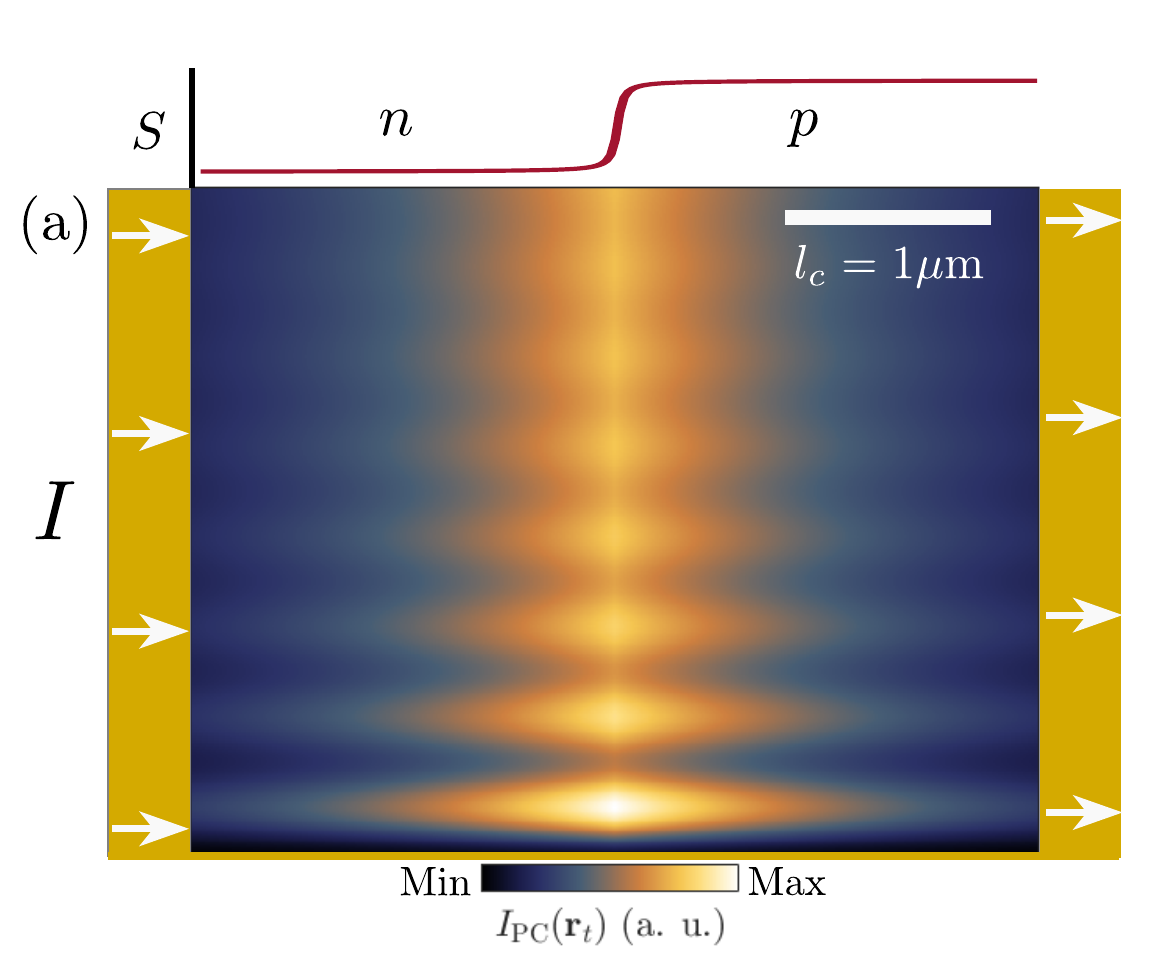}
	\includegraphics[width=3.0 in]{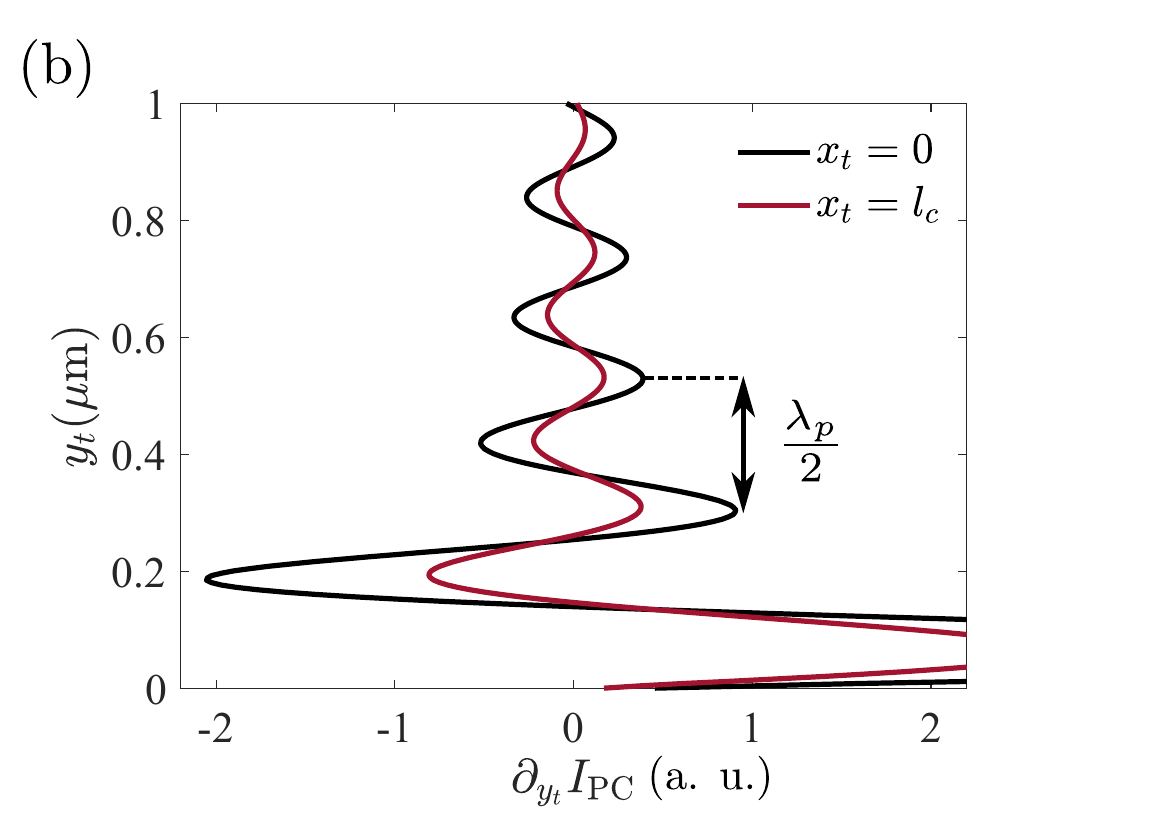}
	\caption{
		(a) 
		The photocurrent as a function of the tip position near a $p$--$n$ junction. 
		The periodic fringes are formed by interference of tip-launched plasmons with their reflections by the sample edge (gold line at the bottom). 
		The scale bar is the cooling length $l_c = 1\, \mu$m.
		The $p$--$n$ junction is modeled as a sharp step-like discontinuity in thermopower (red curve in the top plot). The plasmon wavelength is ${\lambda_p} = 0.2 l_c$ on both sides of the junction, noniniformity of ${\lambda_p}$ at the junction is neglected. 
		The substrate is assumed to be a perfect heat conductor maintaining a constant temperature. 
		(b)
		The linear cuts through panel (a) in the $y$-direction, parallel to the junction, at $x_t = 0$ and $x_t = l_c$. 
	}
\label{fig:Device_Plots}
\end{figure*}

As an illustrative example of how plasmons show up in the PT photocurrent signal, we consider the case
where the sample contains a $p$-$n$ junction along the $y$-axis, i.e., $x = 0$, which is normal to the sample edge at $y = 0$.
This geometry models the case studied experimentally~\cite{AlonsoGonzalez2016}.
We approximate the thermopower profile by a step-like function of coordinate $x$.
We also assume $\sigma^\mathrm{DC}$ is uniform, thereby neglecting the suppression of $\sigma^\mathrm{DC}$ in a neighborhood of the junction.
To obtain the temperature distribution, we again use the method of images approximation:
\begin{equation}
T(\vec{r},\vec{r}_t) = T_t(x - x_{t}, y - y_{t}) \pm T_t(x - x_{t}, y + y_{t}),
\label{eqn:T_GF_image} 
\end{equation}
where $T_t(x, y)$ is the temperature profile in an infinite uniform sample,
the top (bottom) sign corresponds to a boundary with vacuum (metal), and we set the ambient temperature $T_0$ to zero, for simplicity. (The value of $T_0$ does not affect the result for the photocurrent.) 

The spatial distribution of the photocurrent signal computed within this model is shown in Fig.~\ref{fig:Device_Plots}.
The photocurrent exhibits plasmonic fringes as a function of the tip coordinate $y_{t}$ due to the interference of the tip-launched modes with their reflections off the edge,
as observed in the experiment~\cite{AlonsoGonzalez2016, Menabde2021}.
It can be seen from Fig.~\ref{fig:Device_Plots}(a) that appreciable photocurrent is generated only when the
lateral tip position is close enough to the $p$-$n$ junction, $|x_{t} | \lesssim l_c$.
The cooling length $l_c$ is defined in Sec.~\ref{ssec:TR}. It was chosen to be $l_c \approx 1\,\mu\mathrm{m}$ in this calculation. 
The fringe periodicity does not depend on $x_t$; however, the fringe amplitude decreases with $x_t$,
see Fig.~\ref{fig:Device_Plots}(b).
The decay law of the excess temperature  $T_t(x, y)$ (and so, photocurrent) as a function of distance is discussed in more detail in Sec.~\ref{ssec:TR}. 

\begin{figure}[th]
\includegraphics[width=3.0 in]{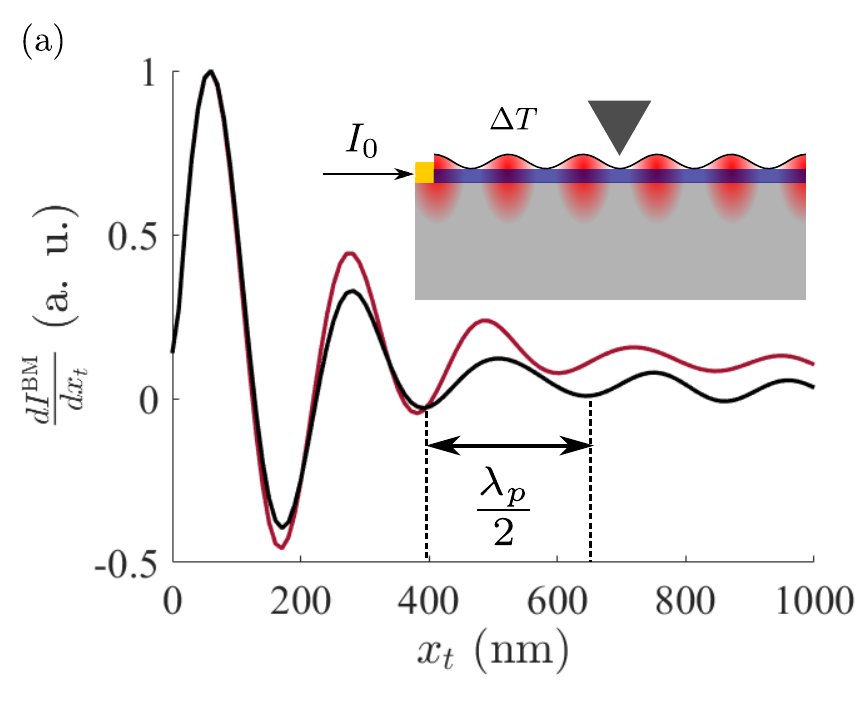}
\includegraphics[width =3.0 in]{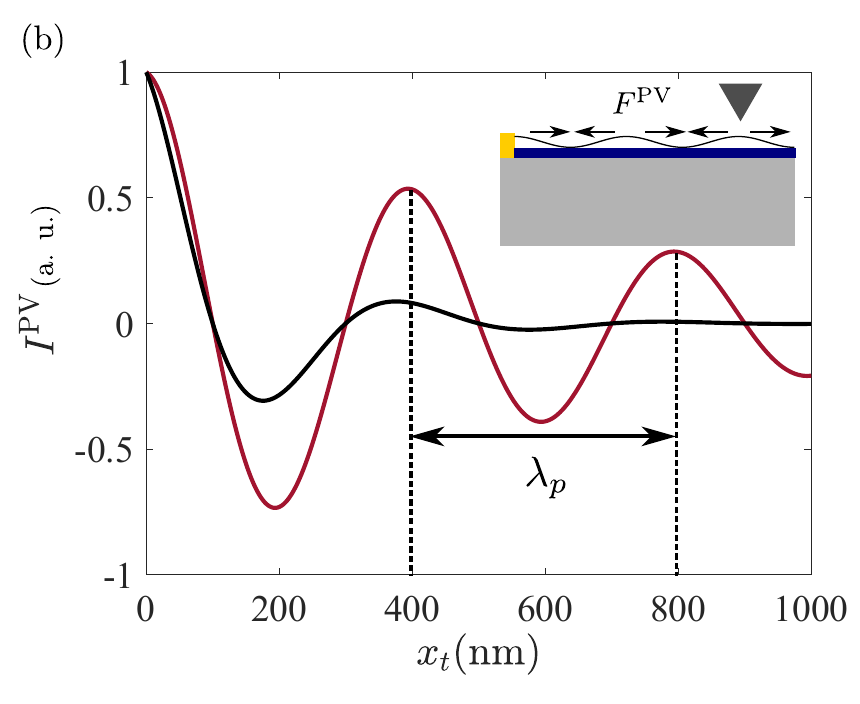}
\caption{
	(a) Spatially periodic photocurrent near a contact due to the BM mechanism. The signal is normalized to its maximum value. 
	Computational parameters are the same as in Fig.~\ref{fig:Device_Plots}, with the red curve illustrating a reduced cooling length $l_c = 400$ nm. 
	Inset: schematic of the sample with temperature oscillations which arise due to the standing wave pattern formed by the field near the edge. The tip is represented by the inverted pyramid. 
	(b) Fringes near a contact of an unbiased sample, due to the PVC. The current is normalized to its maximum.
	The red curve corresponds to a plasmon damping rate of $\omega \tau = 2$, and the black curve corresponds to a damping rate $\omega \tau = 10$. 
	Inset: schematic of the sample geometry, where fringes of period $\lambda_p$ come from the ponderomotive force $F^{\mathrm{PV}}$. }
\label{fig:BMPV_plot}
\end{figure}

\subsubsection{Biased sample}
\label{sssec:biased}
If the sample carries a DC current already in the absence of light,
another possible mechanism of photocurrent plasmonic fringes is the BM effect.
The scanned probe locally modifies the temperature distribution, which causes a local variation in the conductivity.  
The size of the BM effect is proportional to the derivative ${\partial\sigma^\mathrm{DC}} / {\partial T}$,
which depends on details of electron scattering,
see Sec.~\ref{sec:equations}.
The total BM photocurrent in a rectangular sample of size $L \times W$
is
\begin{equation}
I^\mathrm{BM} = -\frac{V}{L W}\, \frac{\partial \sigma^\mathrm{DC}}{\partial T}
\int\limits d^2 r  \, [T(\mathbf{r}) - T_0].
\label{eqn:I_bolo}
\end{equation}
The results of our calculations done using this equation are plotted in Fig.~\ref{fig:BMPV_plot}(a).
They exhibit the now familiar ${\lambda_p} / {2}$ fringes near the contacts. 
It appears that there are many similarities between the spatial features observable in PT and BM photocurrent. 
In fact, $I^\mathrm{BM}$ in Eq.~\eqref{eqn:I_bolo} depends on the tip position
in exactly the same way as the PT photocurrent in a sample with a linearly varying thermopower $S(x, y) \propto x$. 
%
%

In principle, photocurrent in a biased sample can be also generated through the PT because the source-drain bias causes self-gating,
i.e., the carrier density change $\Delta n \sim C V / e$ across the sample, with $C$ being the capacitance. 
To show that the BM should normally be more important than the PT, we compare the coefficients of Eqs.~\eqref{eqn:I_bolo} and~\eqref{eqn:I_SG_n_edge}. 
Using Eq.~\eqref{eqn:Mott} for the thermopower and a rough estimate ${\partial\sigma^\mathrm{DC}} / {\partial T} \sim {\sigma^\mathrm{DC}} / {T}$,
we find
\begin{equation}
\label{eqn:Ibol_Ipte}
\frac{I^\mathrm{PT}}{I^\mathrm{BM}} \sim \frac{T^2}{\mu^2}\, \frac{1}{k_F d_1} \ll 1.
\end{equation}
Indeed, experiments show that the BM contribution typically dominates over the PT one~\cite{Freitag_2012}.

\subsection{Interference fringes due to the plasmons: coherent PV effect}
\label{ssec:PC}

The remaining contribution to the photocurrent introduced in Sec.~\ref{sec:intro} is the PV term.
The PV effect is a complicated phenomenon that depends on many microscopic details of the system. 
Therefore, as our final example, we study the appearance of plasmonic fringes within a simple
representative model.
We assume again that our graphene sample is uniformly doped, $\sigma^\mathrm{DC} = \text{const}$ and has the shape of a $L \times W$ rectangle with contacts at $x = 0$ and $x = L$, as in
Fig.~\ref{fig:heterostructure_schematic}.
In the absence of an applied bias voltage $V$,
the chemical potential and the thermopower are also uniform,
$\mu, S = \text{const}$.
Hence, the BM, PT, and the thermal PV photocurrents all vanish.
What remains is the \textit{coherent} PV current,
which is given by Eq.~\eqref{eqn:jpv}.
A particularly simple result is obtained in the hydrodynamic regime, where the largest contribution to $\sigma_{ilm}^{\mathrm{PT}}$ in the limit $\omega\gg \Gamma_d$ is given by the ponderomotive force $e\vec{F}^{\mathrm{PV}}(\vec{r})$~\cite{Aliev1992,Sun2018}: 
\begin{equation}
\label{eqn:j_PV_ponder}
\vec{j}^\mathrm{PVC}(\vec{r})
= \sigma^\mathrm{DC}\, \vec{F}^{\mathrm{PV}}(\vec{r})\,,
\quad  \vec{F}^{\mathrm{PV}}(\vec{r}) = -\frac{e}{m \omega^2}\, \nabla \left|\vec{E}(\vec{r}, t)\right|^2. \end{equation}
Therefore, the current is proportional to the difference of field intensities at the contacts.
If the field on the contact is the sum of the  external field $\vec{E}\, e^{-i\omega t} + \mathrm{c.c.}$ and the field $\vec{F}(\vec{r})$ created by the tip, the PV current is
\begin{equation}
I^{\mathrm{PVC}}(x_t)\propto \cos(q_p |x - x_t| + \varphi)\bigg|^{x = L}_{x = 0}. 
\label{eqn:I_ptsource}
\end{equation}
Here, $\varphi$ is a phase shift which contains the reflection coefficient from the contact, as well as any phase shift depending on the properties of the tip. 
Equation~\eqref{eqn:I_ptsource} is plotted in Fig.~\ref{fig:BMPV_plot}(b). 
When the tip is located halfway between the source and drain, the photocurrent vanishes, since then the tip-sample system is inversion-symmetric. 
The spatial period of the fringes is $\lambda_p = {2\pi} / {q_p}$, which is twice the period of the standing waves observed in the PT, BM, and in s-SNOM.
This qualitative difference is a result of the interference between the external and the tip-launched field, in contrast the the interference between the launched and reflected waves in PT or s-SNOM,
see Table~\ref{tbl:overview}.
More complicated models would be necessary to accurately model plasmon reflection from the contact~\cite{Rejaei2015} or the effect of probe shape and composition~\cite{Jiang2016} on the photocurrent.

\begin{table}[bth]
	\begin{ruledtabular}
		\begin{tabular}{llcc}
			Effect type & Photocurrent scaling & Fringe period
			\\
			\hline 
			PT & $E^2 \Delta S$  &  $\lambda_p / 2$
			\\[3pt]
			PV & $E^2 \exp(-x_t / L)$  &  $\lambda_p$
			\\[3pt]
			BM & $E^2 E^{\mathrm{DC}}$  &  $\lambda_p / 2$
		\end{tabular}
	\end{ruledtabular}
	\caption{
		Summary of the studied phototocurrent mechanisms producing interference fringes.
	}
	\label{tbl:overview} 
\end{table}

\section{Photocurrent spectra versus \lowercase{s}-SNOM spectra}
\label{sec:frequency} 

In the preceeding Section we dealt with the tip-position dependence of the photocurrent.
Collective mode effects have also been observed in the frequency dependence of
both photocurrent nanoscopy~\cite{Koppens_2014} and the s-SNOM measurements~\cite{Fei2011}. 
It is interesting to ask whether the frequency dependencies obtained through these two techniques can be related to one another. 
Below we discuss one case where such a mathematical relation can be established.

The s-SNOM signal is a measure of the dipole moment induced on the probe by the sample ~\cite{Keilmann2004}.
For a long and thin probe oriented normal to the surface, this signal is proportional to the out-of-plane component of the probe dipole moment $p^z$: 
\begin{equation}
S_{\mathrm{SNOM}}(\omega) = C_1 p^z(\omega)
\label{eqn:PC_SNOMcorrespondence}
\end{equation}
with some frequency-independent constant of proportionality $C_1$. 
The total power dissipated by the probe driven by an external field $\vec{E}_{\text{ext}} = \vec{E}_{0}e^{-i\omega t} + \mathrm{c.c.}$ can also be expressed through its dipole moment:
\begin{equation}
\label{eqn: J_heat_tot}
P_{\mathrm{tot}}(\omega) =  \omega \im \vec{p}(\omega)\cdot\vec{E}_{\text{ext}}.
\end{equation}
PT, BM, and PVT photocurrents depend on the excess temperature, which is proportional to the Joule heating $T - T_0 \propto |E|^2$. 
In fact, we can show that for
the setup examined in Sec.~\ref{ssec:PR}, i.e., the tip near a domain wall in TBG, the PT photocurrent is proportional to the total Joule heating: 
\begin{equation}
\label{eqn:I_pc_far}
S_{\mathrm{PC}}(\omega) = C_2 P(\omega), \quad  P(\omega) = \int\! d^2 r p(\vec{r},\omega)\,, 
\end{equation}
see Eqs.~\eqref{eqn:Joule_H}, \eqref{eqn:T_GF_near}, and \eqref{eqn:I_PC_2D}.
As with $C_1$, constant $C_2$ has no frequency dependence.  

Within the quasistatic approximation, radiation losses are negligible, so the dissipation is dominated by losses in the sample. 
Furthermore, if dielectric losses are negligible compared to the Joule heating, $P_{\mathrm{tot}}(\omega) \approx P(\omega)$ by conservation of energy. 
The equality of losses in the sample and the dissipation by the polarizable probe provides an elegant connection between the s-SNOM signal $S_{\mathrm{SNOM}}$ and photocurrent $S_{\mathrm{PC}}$:
\begin{equation}
\label{eqn:PC_Snom_connection}
S_{\mathrm{PC}}(\omega) \propto \omega \im S_\mathrm{SNOM}(\omega). 
\end{equation}
Although we worked within the simple point-dipole approximation in Sec.~\ref{sec:spatial}, this correspondence only requires Eq.~\eqref{eqn:PC_SNOMcorrespondence} to be applicable, allowing more complicated models developed for s-SNOM modeling~\cite{Cvitkovic2007, McLeod2014, Jiang2016} to be extended to near-field photocurrent techniques.

Finite dielectric losses lead to deviations
from Eq.~\eqref{eqn:PC_SNOMcorrespondence} and so Eq.~\eqref{eqn:PC_Snom_connection}.
A rough validity condition for Eq.~\eqref{eqn:PC_SNOMcorrespondence} can be obtained from the point dipole model of the tip~\cite{Cvitkovic2007}. It is given by
\begin{equation}
\im \varepsilon(\omega) \ll
\frac{4\pi \varepsilon_0}{\omega a}\, \re \sigma(\omega) , 
\label{eqn:PC_SNOM_wcondition}
\end{equation}
where $a$ is the radius of curvature of the tip and $\varepsilon$ is the permittivity of the substrate. 

\begin{figure}[t]
\includegraphics[width=2.6in]{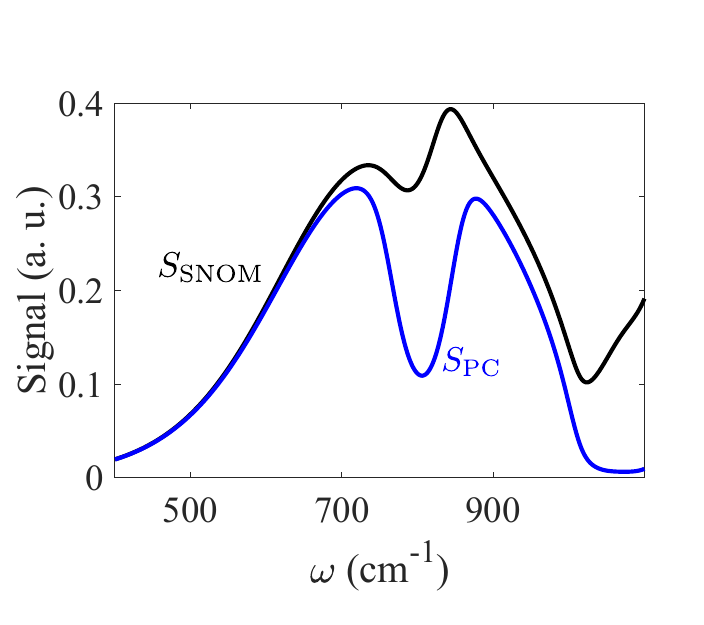}
\caption{A comparison of the total power dissipated by the heterostructure (upper curve) and the in the graphene layer alone (lower curve), normalized to their respective maxima.
The former is proportional to the imaginary part of the s{\-}SNOM signal $S_\mathrm{SNOM}$, and the latter is proportional to the photocurrent $I^\text{PT}$. 
The plasmon resonance in graphene results in a broad peak in both signals at $\omega \sim 700 \, \mathrm{cm}^{-1}$
for the Fermi energy of graphene $\mu^\text{DC} = 1800\, \mathrm{cm}^{-1}$. 
The dip in the photocurrent occurs at the phonon resonance 
$\omega = 800 \, \mathrm{cm}^{-1}$ of the SiO$_2$ substrate.
}
\label{fig:SNOM_G}
\end{figure}

To illustrate how well Eq.~\eqref{eqn:PC_SNOMcorrespondence} works in a concrete example, we consider a system where graphene layer is deposited on SiO$_2$ substrate near a $p$-$n$ junction.
The modeled s-SNOM signal, PT photocurrent, and their ratio are illustrated in Fig.~\ref{fig:SNOM_G}.  
The peak in the signal due to the graphene plasmon, observed both in s-SNOM~\cite{Fei2011} and photocurrent~\cite{Freitag2013}, is present in both measurements.
This sample satisfies the low-loss condition Eq.~\eqref{eqn:PC_SNOM_wcondition} away from the phonon resonances of SiO$_2$ that occur near $\omega = 800\, \mathrm{cm}^{-1}$ and $1120\, \mathrm{cm}^{-1}$~\cite{Kucirkova1994}.
(The latter is not visible in Fig.~\ref{fig:SNOM_G}.)

The above discussion suggests that a photocurrent signal relates more directly to the properties of the 2D conductor and less to those of the substrate, potentially making photocurrent measurements more suitable for extracting properties of conducting layers embedded in complicated heterostructures. 

\section{Model}
\label{sec:equations}

\subsection{Optical response of a layered medium}
\label{ssec:OR}

In this Section we go over computational aspects of our modeling.
We start with presenting the set of equations we have used to compute the optical response of
the systems discussed in the previous Section.
All those model systems are layered heterostructures of the type depicted in Fig.~\ref{fig:heterostructure_schematic}.
We number the layers sequentially top to bottom.
The vacuum half-space above the sample is layer $0$.
The bottom substrate, which we also treat as semi-infinite, is layer $M \geq 1$.
We allow for a uniaxial anisotropy of the layer materials, such that
the in- and out- of plane permittivities $\varepsilon_{m}^{\perp, z}$ of layer $m$
may be unequal.
Additionally, if any of the constituent materials can be considered 2D, we do not assign it an index.
Instead, we model it as a zero-thickness sheet of ac conductivity $\sigma_{mn}(\omega)$ at the interface of layers $m$ and $n = m + 1$.
(If no such 2D material is present at that interface, then $\sigma_{mn} = 0$.)

In general, the optical response of the system is determined by the
reflection coefficients $r^\alpha$ of polarizations $\alpha = p$ or $s$. 
However, in the near-field limit, only the
$p$-polarization reflection coefficient $r^p = r^p(q, \omega)$ is important.
This quantity can be computed from the following recursion formula~\cite{Wu2015}: 
\begin{align}
	\label{eqn:rP_recursion}
	&r_j = r^p_{j, j+1} - \frac{(1 - r^p_{j, j + 1})(1 - r^p_{j+1, j}) r_{j+1}}{r^p_{j + 1, j} r_{j + 1} - \exp(-2 i k^z_{j+1} d_{j+1})},\\
	& k^z_{j+1} = \sqrt{\varepsilon^\perp_{j+1}}\, \sqrt{\frac{\omega^2}{c^2} - \frac{q^2}{\varepsilon^z_{j+1}}}\,,
\end{align}
where $q$ is the in-plane momentum,
$d_{j}$ is the thickness of layer $j$, and $r^{p}_{n m}$ is the reflection coefficient
of the interface between layers $n$ and $m$:
\begin{equation}
	\label{eqn:rP_def_QS}
	r^{p}_{n m}(q, \omega) = \frac{\varepsilon_m - \varepsilon_n + \frac{4\pi i\sigma_{m n} q}{\omega}}{\varepsilon_m + \varepsilon_n + \frac{4\pi  i  \sigma_{m n} q}{\omega}}, \quad \varepsilon_m = \sqrt{\varepsilon_m^\perp}\, \sqrt{\varepsilon_m^z}.
\end{equation}
The recursion starts with $j = M - 1$ for which $r_{M - 1} = r^p_{M - 1, M}$, and continues to progressively smaller $j$.
The reflection coefficient of the entire system is given by $r_0$.
For real $q < \omega / c$, the reflection coefficient has an absolute value smaller than unity.
Away from this radiative zone, function $r^p(q, \omega)$ may have poles at some complex $q$ that
have relatively small imaginary parts.
Such poles define the dispersion of the propagating collective modes of the system whose effect on photocurrent we want to study.

\begin{figure}[th]
	\includegraphics[width=3.10 in]{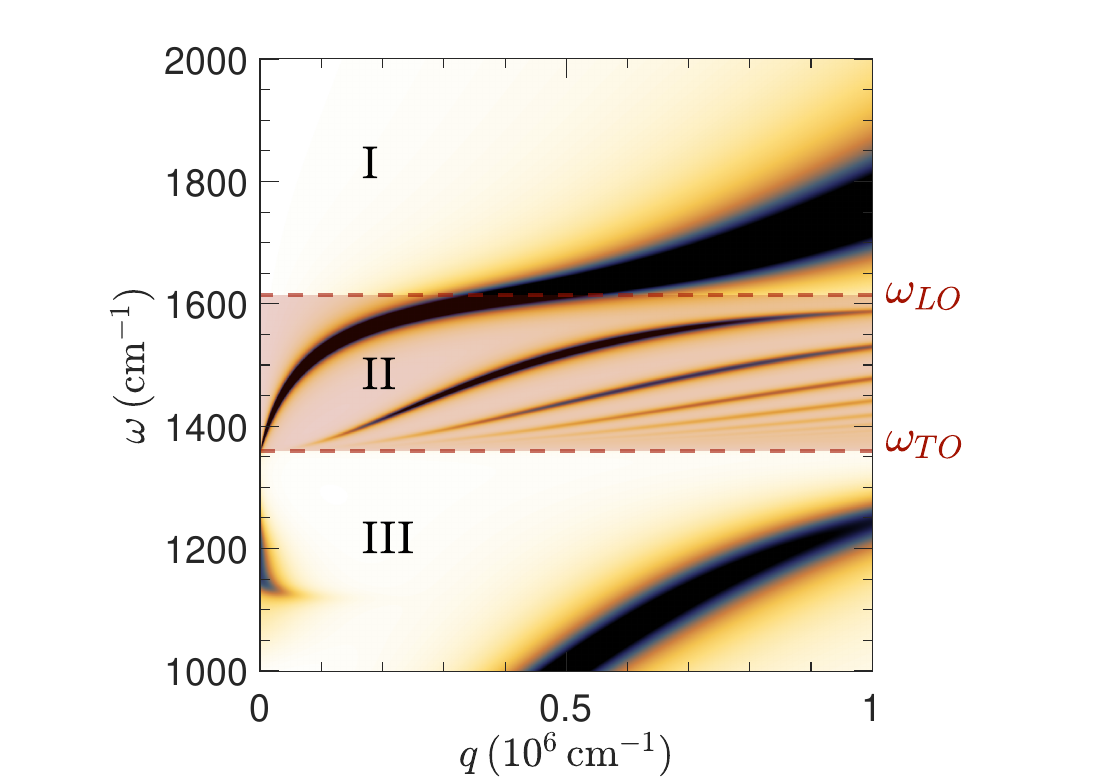}
	\caption{The reflection coefficient $\im r^p(q,\omega)$ of a  graphene-hBN-SiO$_2$ heterostructure. 
	The maxima in this pseudocolor plot correspond to the dispersions of hybridized collective modes known as the plasmon phonon polaritons. 
	The modes in Regions I and III are plasmon-like. 
	Region II, where hBN acts as a hyperbolic optical medium, contains multiple dispersion lines of waveguide polariton modes.
	These waveguide modes exhibit avoided crossings with the graphene plasmon. 
	The small peak near $\omega = 1170\, \mathrm{cm}^{-1}$ in Region III is due to the phonon mode of the hBN/SiO$_2$ interface.  
	The graphene chemical potential is $\mu = 2400\, \mathrm{cm}^{-1}$, the hBN thickness is $d = 50\, \mathrm{nm}$.}
	\label{fig:ColModes}
\end{figure}

An illustrative example of function $r^p(q, \omega)$ is shown in Fig.~\ref{fig:ColModes}.
It is computed for a heterostucture consisting of a doped monolayer graphene placed on a $50\unit{nm}$-thick hBN crystal, which is in turn placed on a bulk SiO$_2$ substrate.
At low frequencies $\omega < \omega_{\mathrm{TO}}$
(Region I) the dispersion of this system contains a single branch,
which is basically the plasmon mode of graphene.
However, there is also a weak feature present near $\omega = 1100\, \mathrm{cm}^{-1}$,
which is due to the interface phonon of hBN and SiO$_2$.
In a range of intermediate frequencies
$\omega_{\mathrm{TO}} < \omega < \omega_{\mathrm{LO}}$
(Region II) 
where hBN acts as a hyperbolic material,
with $\re\varepsilon^{\perp}(\omega) < 0 < \re\varepsilon^{z}(\omega)$, 
there are multiple dispersion branches.
These are known as hyperbolic phonon polaritons.
More precisely, 
these modes result from hybridization of the graphene plasmon with phonon polaritons of hBN, and so they should be referred to as the hyperbolic plasmon phonon polaritons.
At high frequencies $\omega > \omega_{\mathrm{LO}}$
there is only a single plasmon branch.

Our goal in this paper is to understand the effect of these collective modes on photocurrent measured by scanned probes.
Let us now discuss the electric field produced by such a probe.
Computing this field from a realistic model can be quite laborious.
Instead, as common in the s-SNOM literature~\cite{Keilmann2004, Cvitkovic2007, Fei2011, Jiang2016},
we model the tip of the probe by a polarizable dipole of amplitude $p^z \hat{\vec{z}}$ positioned a distance $z_t$ from the sample.
If $z_t \ll c / \omega$,
the field produced by the tip can be computed within the
quasi-static approximation.
The scalar potential produced by the tip in the half-space $z \geq 0$ above the sample is the superposition of the bare and the reflected dipole potentials:
\begin{align}
	&\tilde{\Phi}(\vec{q}, z) = \tilde{\varphi}(\vec{q}, |z - z_t|) - r^p(\vec{q},\omega)\tilde{\varphi}(\vec{q}, z + z_\text{t}),
	\label{eqn:Phi_FS}\\
	& \tilde{\varphi}(\vec{q}, z) = \frac{2\pi p_z}{\varepsilon_0}\,
	 e^{-q |z|}.
	\label{eqn:V_FS}
\end{align}
In turn, the in-plane field $\vec{E}(\vec{r}) = -\nabla\Phi(\vec{r}, 0)$ 
and the local Joule heating $p(\vec{r})$ in the 2D layer where the photocurrent is produced are given by
\begin{align}
	&\vec{E}(\vec{r}) = \hat{\vec{r}} \int\limits_0^\infty \frac{dq}{2\pi}   q^2 \tilde{\Phi}(q, 0) J_1(q r),
\label{eqn:E_inplane}\\
	&p(\vec{r}) = \frac{1}{2} \re \sigma(\vec{r}, \omega)\, |\vec{E}(\vec{r})|^2,
\label{eqn:Joule_H}
\end{align}
where $J_\nu(z)$ is the Bessel function of the first kind of order $\nu$.
Here we assume that the AC conductivity $\sigma(\vec{r}, \omega)$
of the 2D layer varies slowly on the scale of $z_t$.

\subsection{Plasmonic response of the 2D layer}
\label{ssec:CCR}

The continuity equation for a time-harmonic perturbation of frequency $\omega$ to the charge density $\rho = en $ is
\begin{equation}
	-i \omega \rho + \partial_i j_i  = 0\,.
	\label{eqn:Charge_cont}
\end{equation}
The collective excitations of the system, e.g., plasmon-phonon modes illustrated by Fig.~\ref{fig:ColModes} invariably involve oscillations of $\vec{j}$ and $\rho$.
We can center our attention on the 2D layer and consider the rest of the system an environment.
It is then possible to take a point of view that
all collective excitations are 2D plasmons renormalized by the environment.
Within this approach, the derivation of the mode spectra goes as follows. 
First, we find the reflection coefficient $r_p^*(q, \omega)$ of the system without the conducting 2D layer on top, by the procedure explained above.
All the charges above the sample are now considered the sources of an external potential
$\Phi_\mathrm{ext}$, which is computed similar to
Eq.~\eqref{eqn:Phi_FS} except with $r^p(q, \omega)$ replaced by $r_p^*(q, \omega)$.
The total in-plane potential $\Phi = \Phi_\mathrm{ind} + \Phi_\mathrm{ext}$ is the sum of this $\Phi_\mathrm{ext}$ and
the potential $\Phi_\mathrm{ind}$ induced by the 2D layer's own charge $\rho$:
\begin{equation}
	\tilde\Phi_\mathrm{ind}(\vec{q}, \omega) =
	\frac{2\pi}{\varepsilon_0 q}[1 - r_p^*(q, \omega)] \tilde\rho(\vec{q}, \omega)\,.
	\label{eqn:Gauss}
\end{equation}
(Here we again assume the quasi-static limit $q \gg \omega / c$.)
Combining these equations, we arrive at
\begin{equation}
	\tilde\Phi(\vec{q}, \omega) = {\tilde\Phi_\mathrm{ext}(\vec{q}, \omega)} / {\varepsilon_\mathrm{2D}(q,\omega)}\,,
	\label{eqn:Phi_Phi_Phi}
\end{equation}
where the function
\begin{equation}
	\varepsilon_\mathrm{2D}(q, \omega) = 1 - 
	\frac{1 - r_p^*(q_p, \omega)}{\varepsilon_0}\,
	\frac{2\pi \sigma(\omega) q}{i\omega} 
	\label{eqn:epsilon_2D}
\end{equation}
has the physical meaning of the effective 2D permittivity of the conducting layer.
The relation between $r_p$, $r_p^*$, and $\varepsilon_\mathrm{2D}$ is
\begin{equation}
	1 - r_p = \frac{1 - r_p^*}{\varepsilon_\mathrm{2D}}\,.
	\label{eqn:r_p^*}
\end{equation}
The imaginary part of $r_p$ characterizes the losses of the system. 
For a general multilayer structure, the low-loss condition can be expressed as
\begin{equation}
	\im r_p(\bar{q}, \omega) \gg \im \{\varepsilon_{2D} \left[r_p^*(\bar{q}, \omega)-1\right]\},
\end{equation}
at $\bar{q}\sim a^{-1}$, where $a$ is the radius of curvature of the near-field probe, see Eqs.~\eqref{eqn:rP_recursion} and \eqref{eqn:Gauss}.
For a graphene layer at the interface of two semi-infinite media, one recovers Eq.~\eqref{eqn:PC_SNOM_wcondition}. 
The sought mode dispersions are the poles of $r_p$ or equivalently, the zeros of $\varepsilon_\mathrm{2D}(q, \omega)$.
At a given $\omega$, these zeros occur at momenta $q_p$ that solve the equation
\begin{equation}
	q_p = \frac{\varepsilon_0}{1 - r_p^*(q_p, \omega)}\,
	 \frac{i\omega}{2\pi \sigma(\omega)}\,.
\label{eqn:q_plasmon0}
\end{equation}
In general, such $q_p$ are complex and the corresponding collective modes are well defined (underdamped) only if $\im q_p \ll \re q_p$.
Within the Drude model [Eq.~\eqref{eqn:sigma_AC}], 
$\im \sigma / \re \sigma = {\omega} / {\Gamma_d}$,
so the necessary condition for underdamped plasmons to exist is $\omega \gg \Gamma_d$.

The potential of a plasma wave launched by a local source (such as an s-SNOM tip) is given by
\begin{align}
	\Phi(\vec{r}) &\simeq V_t H_0^{(1)}(q_p r)\,,
	\label{eqn:circular_wave}\\
	V_t &\propto \int q_p^2 \Phi_{\mathrm{ext}}(\vec{r}) d^2 r\,,
	\label{eqn:V_t}
\end{align}
where $H_0^{(1)}(z)$ is the Hankel function of the first kind.
For an arbitrary sample-gate separation $d_1$, Eq.~\eqref{eqn:circular_wave} remains universally valid
in the range of distances $z_t \ll r \ll (\im q_p)^{-1}$ but Eq.~\eqref{eqn:V_t} may be modified.
If such distances play the dominant role in the photocurrent response
and the absolute magnitude of this response is not of primary interest,
then it is permissible to use the simple equations~\eqref{eqn:circular_wave}--\eqref{eqn:V_t} to find the potential $\Phi(\vec{r})$, see Sec.~\ref{ssec:PC}.

If the 2D layer resides on a hyperbolic film of thickness $d_1$, e.g., if $\varepsilon_1^\perp < 0$, $\varepsilon_1^z > 0$,
then $r_p^*(q, \omega)$  is given by 
\begin{equation}
	r_p^* = \frac{r_{01}e^{i q \delta} - r_{21} 
	}{e^{i q \delta} - r_{01} r_{21}}.
	\label{eqn:r_P_dipole}
\end{equation}
In this case, Eq.~\eqref{eqn:q_plasmon0} has an infinite number of roots, each representing a different plasmon-phonon eigenmode, see region II in Fig.~\ref{fig:ColModes}. 
The solutions are separated by the constant value
\begin{equation}
\Delta q = \frac{i \pi}{d_1}\, \frac{\sqrt{\varepsilon_1^z}}{\sqrt{\varepsilon_1^\perp}}\,. 
\label{eqn:Delta_q}
\end{equation}
We can use Eq.~\eqref{eqn:r_P_dipole} to find the in-plane components of the electric field: 
\begin{equation}
	\begin{split}
		\qquad &E(r, z_t) =  3 p_z (1 - r_{01}) \ \times \\ &\left[
		e_0(r) + (1 + r_{01}) r_{21} \sum_{k = 1}^{\infty} (r_{01} r_{21})^{k - 1} e_k(r)
		\right],
		\label{eqn:E_r_images}
	\end{split}
\end{equation}
where
\begin{equation}
	e_{k}(r) = \frac{(-i z_t + k \delta) r}{\left[\left(-i z_t + k \delta\right)^2 - r^2\right]^{5/2}} \,.
	\label{eqn:e_n}
\end{equation}
Since $|1 + r_{01}|>1$ in the hyperbolic regime, the largest term in the series is the $k=1$ term. 
These fields have maxima at the concentric rings of radius $r_k$, given by Eq.~\eqref{eqn:r_n}. 
If the slab were made of a non-hyperbolic material, then $\delta$ would be imaginary and the only real roots of Eq.~\eqref{eqn:r_n} would be $r_0 = \frac{z_t}{2}$.
For instance, in an isotropic material $\delta = 2 i d$ and the rings are absent. 
The effect of $e_k$ for large $k$ is negligible in this case, since the image dipoles become progressively further from the origin.

\subsection{Second-order response of a 2D layer}
\label{ssec:NSO}

If the electric field is not too strong, the response of the system can be studied
by expanding all quantities of interest in power series of the electric field.
For example, the current density has the expansion $\vec{j} = \vec{j}^{(1)} + \vec{j}^{(2)} + \ldots$
The second term, quadratic in $E$, can be expressed in terms of the second-order nonlinear conductivity
$\sigma_{ilm}^{(2)}\left(\vec{k}_1, \omega_1;\vec{k}_2, \omega_2\right)$ entering
Eq.~\eqref{eqn:sigma2split}.
For the DC photocurrent generated by a monochomatic field $\vec{E}(\vec{r}) e^{-i \omega t} + \mathrm{c.c.}$ of frequency $\omega$,
the parameter choice $\omega_2 = -\omega_1 = \omega$ is appropriate,
such that
\begin{align}
	j_i^{(2)}\left(\vec{r}\right) &=
	\int \frac{d^2\! {k}_1 d^2\! {k}_2}{(2\pi)^4}\,
	\sigma_{ilm}^{(2)} (\vec{k}_1, -\omega;
	\vec{k}_2, \omega)
	\notag \\
	&\times \tilde{E}_l^* (-\vec{k}_1) \tilde{E}_m(\vec{k}_2)  e^{i (\vec{k}_1 + \vec{k}_2) \cdot \vec{r}},
	\label{eqn:j_from_sigma^2}
\end{align}
where
\begin{equation}
	\tilde{\vec{E}}(\vec{k}) \equiv
	\int d^2 r  e^{-i \vec{k} \cdot \vec{r}}\, \vec{E}(\vec{r})\,.
	\label{eqn:tildeE}
\end{equation}
The functional form of $\sigma_{ilm}^{(2)}$ is highly system-dependent. 
One particular case attracting much interest recently is where the electrons behave collectively, as a fluid~\cite{Lucas2018}. 
This regime is realized when the momentum-conserving electron-electron scattering rate $\Gamma_{ee}$ exceeds the momentum relaxation rate $\Gamma_d$. 
In this hydrodynamic regime, the derivation of the second-order non-linear response simplifies greatly. 
We summarize it in Appendix~\ref{appendix:HD_equations}. 

The first-order ac conductivity is given by the Drude formula
\begin{equation}
		\sigma(\omega) = \frac{\Gamma_d}{\Gamma_d - i \omega}\,  \sigma^\mathrm{DC}\,,
		\quad
		\sigma^{\mathrm{DC}} = \frac{e^2 n}{m}\, \frac{1}{\Gamma_d}\,.
		\label{eqn:sigma_AC}
	\end{equation}
For the second-order current, we find the following combination of terms:
\begin{equation}
	\vec{j}^{(2)} = \vec{j}^\mathrm{PT} + \vec{j}^\mathrm{PVT} + \vec{j}^\mathrm{PVC}\,.
	\label{eqn:j^2}
\end{equation}
The first term is the PT current
\begin{equation}
	\vec{j}^\mathrm{PT} = -\sigma^\mathrm{DC}\, \frac{s}{e n}\, \nabla T.
	\label{eqn:j_PT}
\end{equation}
Comparing with Eq.~\eqref{eqn:jPT}, we see that the thermopower coefficient $S$ is equal to the entropy per unit charge
\begin{equation}
	S = \frac{s}{e n} = \frac{\pi^2}{3}\, \frac{k_B^2 T}{e n}
	\left(\frac{\partial \mu}{\partial n}\right)_{T}^{-1} .
	\label{eqn:Mott}
\end{equation}
The last equation is Mott's formula for a degenerate Fermi gas with a constant scattering rate $\Gamma_d$.
The next term in Eq.~\eqref{eqn:j^2} is the thermal PV current
\begin{equation}
	\vec{j}^\mathrm{PVT} =  -e D \nabla n
	- \frac{\sigma^\mathrm{DC}}{e} \left(\frac{\partial \mu}{\partial T}\right)_n \nabla T\,.
	\label{eqn:j_PVT}
\end{equation}
Lastly, the third term  in Eq.~\eqref{eqn:j^2} is what we previously called the coherent part of the PV current.
After a lengthy but straightforward derivation, one finds  (see, e.g.,~\cite{Sun2018})
\begin{equation}
\begin{split}
	\vec{j}^\mathrm{PVC} &= -\frac{e^3 n}{m^2}\, \frac{1}{\Gamma_d^2 + \omega^2}
	\\
	&\times \left\{
	\frac{1}{\Gamma_d}\,  \nabla|\vec{E}|^2
	 - \frac{2}{\omega}\,
	  \im \left[\vec{E}^*\!\! \left(\nabla \cdot  \vec{E}\right)
	+  \left(\vec{E}^*\!\! \times  \vec{E}\right) \right]
	\right\} .
\end{split}
\label{eqn:j_PVC}
\end{equation}
This expression is well known in plasma physics where it is attributed to the ponderomotive force~\cite{Aliev1992}.
For $\omega\gg \Gamma_d$, the largest term is the first term in the curly brackets, which is Eq.~\eqref{eqn:j_PV_ponder}. 

In the regime of our primary interest $\omega\gg \Gamma_d$, the results for the PV and PT parts of $\sigma_{ilm}^{(2)}$ simplify to
\begin{align}
\sigma_{ilm}^{\mathrm{X}}\left(\vec{k}_1, \omega; \vec{k}_2, -\omega\right) &\simeq -i c^{\mathrm{X}}\, \frac{e^3 n}{m^2 \omega^2}\,
(k_{1i} + k_{2i}) \delta_{lm}\,,
\label{eqn:sigma^X}\\
c^{\mathrm{PV}} &=
	\frac{1}{2 \Gamma_d} + \frac{1}{\Gamma_E} \left(\frac{\partial\mu}{\partial T}\right)_n,
\label{eqn:c_PV}\\
c^{\mathrm{PT}} &= \frac{1}{\Gamma_E}\,.
\label{eqn:c_PT}
\end{align}
Under the assumption $\Gamma_E \ll \Gamma_d$ made earlier, the PT component is the dominant one.
Note that the extra factor of $2$ in the first term of Eq.~\eqref{eqn:c_PV} compared to Eq.~\eqref{eqn:j_PVC} appears when we make the transition from the real to the Fourier space.
In addition, the derivative ${\partial\mu} / {\partial T} \ll 1$ entering the second term in Eq.~\eqref{eqn:c_PV} is very small for a degenerate 2D Fermi liquid $T \ll |\mu|$. 

In the scenario considered in Sec.~\ref{ssec:PC}, only the PVC contributes to the photocurrent. 
Assuming $\omega\gg\Gamma_d$, the current through a resistor $R$ can be obtained from integrating Eq.~\eqref{eqn:j_PVC}: 
\begin{equation}
	I^\mathrm{PVC} 
	= -\frac{R}{R + R_{g}}\,
	\frac{e \sigma^\mathrm{DC}}{m \omega^2 L} \int \partial_x \left| \vec{E}(\vec{r}, t) \right|^2 d^2 r,
	\label{eqn:I_PV_ponder1}
\end{equation}
where $R_g = {L} / {(W\sigma^\mathrm{DC})}$ is the resistance of the sample. 
This integral immediately simplifies to
\begin{equation}
	\label{eqn:I_PV_ponder}
	I^\mathrm{PVC}
	= -\frac{R}{R + R_{g}}\,
	\frac{e \sigma^\mathrm{DC}}{m \omega^2 L} \int\limits_0^{W} d y \left| [\vec{E}(\vec{r}, t)]^2 \right|^{x = L}_{x = 0},
\end{equation}
which indicates that $I^\mathrm{PVC}$ is determined solely by the electric field $E_x$ at the contacts.
In Eq.~\eqref{eqn:I_PV_ponder} we took into account that the contacts are equipotential so that $E_y = -\partial_y \Phi$ must vanish.
The tip-dependent contribution to the field comes from the interference between the tip-generated and external fields: 
\begin{equation}
	I^\mathrm{PVC} \simeq -\frac{2R}{R + R_g}\,
	\frac{e\sigma^\mathrm{DC}E_0}{m\omega^2 L}
	\mathrm{Re}\left(V_t e^{i q_p |x - x_{t}|}\right)\bigg|^{x = L}_{x = 0}, 
	\label{eqn:I_ptsource_detailed}
\end{equation} 
which is Eq.~\eqref{eqn:I_ptsource} with $\varphi = \arg(V_t)$. 
\subsection{Thermal response}
\label{ssec:TR}
\begin{figure}[h]
	\includegraphics[width = 1.3in]{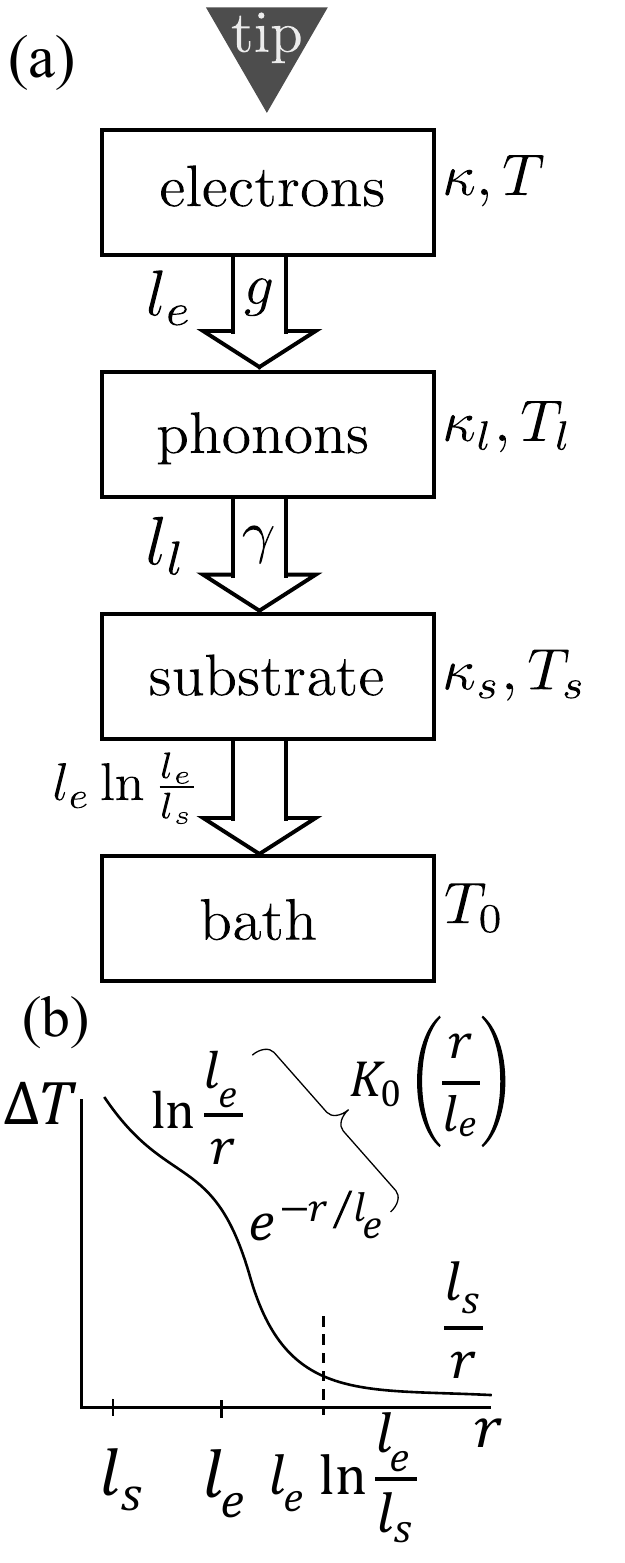}
	\caption{(a) A schematic of the tip-generated heating. 
	Heat from the electrons at temperature $T$ is transferred to the lattice with a mean free path of $l_e$.
	The heat from the lattice at temperature $T_l$ is then transferred to the substrate over a phonon mean free path $l_l$. 
	The heat diffuses through the three-dimensional substrate with temperature $T_s$, which is coupled to a heat bath at a fixed temperature $T_0$. 
	(b) A sketch of the temperature profile produced by a point source for $l_e\gg l_l$.}
	\label{Fig:Appendix_schematic}. 
\end{figure}

The energy relaxation of electrons involves
their interaction with multiple degrees of freedom such as the phonons of the 2D layer and the substrate. 
In this section, we consider a model where we assign separate temperatures to these two subsystems. 
This can be a reasonable approximation if far-from-equilibrium effects (phonon wind, phonon amplification, hot electrons, \textit{etc}.) can be neglected~\cite{Gurevich1989, Andersen2019,Massicotte2021}. 
We introduce the electron-phonon coupling constant $g$ and the inverse of the Kapitza resistance of the graphene-substrate interface $\gamma$ and
write the following three-temperature heat transfer equations:
\begin{align}
&-\kappa \nabla^2 T + g(T - T_{l}) = p(\vec{r}),
\label{eqn:T1}\\
		&-\kappa_{l}\nabla^2 T_{l} + \gamma (T_{l} - T_{s}|_{z = 0}) 
+ g (T_{l} - T) = 0,
\label{eqn:T2}\\
& -\kappa_{s}\left( \nabla^2 + \partial_z^2\right) T_{s} = 0.
\label{eqn:T3}
\end{align}
The Joule heating power $p(\vec{r})$ produced by the scanned probe 
acts as a localized heat source, as shown in Fig.~\ref{Fig:Appendix_schematic}(a).
The electronic and lattice thermal conductivities, $\kappa$ and $\kappa_{l} h$, define three characteristic lengths
\begin{align}
 l_{e} &= \sqrt{\frac{\kappa}{g}}\,,
\label{eqn:l_e}
 \\
 l_l &= \sqrt{\frac{\kappa_{l}}{\gamma + g}}\,,
\label{eqn:l_l}
 \\
l_s &= \frac{\kappa_s}{\gamma}\,. 
\label{eqn:l_s}
\end{align} 
Depending on their values relative to each other and the distance $r$ from the source,
qualitatively different scaling laws for the excess electron temperature $\Delta T(\vec{r}) = T(\vec{r}) - T_0$ emerge, as discussed below. 
For defineness, we consider the case where
electron cooling into the substrate heat sink is
efficiently mediated by graphene phonons: $\kappa_l \gg \kappa$ and $\gamma \gg g$.
For graphene on SiO$_2$,
these assumptions can be justified using
the following parameter estimates~\cite{Freitag2009, Pop2012, Zhu2018}
\begin{align}
\kappa_l &= \kappa_{3\mathrm{D}} h
 = 0.34 \times 10^{-6}\, \frac{\mathrm{W}}{\mathrm{K}}\,,
\label{kappa_l_from_kappa_3D}
\\
\gamma^{-1} &= 4.2 \times 10^{-8}\, \frac{\mathrm{m}^2 \mathrm{K}}{\mathrm{W}}\,,
\quad
\kappa_s = 1.0\, \frac{\mathrm{W}}{\mathrm{m}\,\mathrm{K}}\,.
\label{eqn:kappa_estimates}
\end{align}
In Eq.~\eqref{kappa_l_from_kappa_3D} we employ the commonly reported ``bulk'' thermal conductivity $\kappa_{3\mathrm{D}}$, which is calculated by modeling graphene as a thin film of thickness $h = 0.335\, \mathrm{nm}$.
The phonon mean-free path $l_{ph}$ corresponding to the chosen value
$\kappa_{3\mathrm{D}}
= 1000\, \mathrm{W}\, \mathrm{m}^{-1} \mathrm{K}^{-1}$
is~\cite{Pop2012} $l_{ph} \approx 200\,\mathrm{nm}$.
These parameter values yield the estimate $l_l \simeq (\gamma^{-1} \kappa_{l})^{1 / 2} = 120\, \mathrm{nm}$,
which is substantially shorter
than the typical values of the cooling length reported for encapsulated graphene systems. As discussed below, this is due to the fact that the limiting step
in the cooling process is the energy exchange between electrons and phonons of graphene.
 
To get the remaining parameters in Eqs.~\eqref{eqn:l_e}--\eqref{eqn:l_s},
we first use the Widemann-Franz law
\begin{equation}
	\kappa = \frac{\pi^2}{3 e^2}\, T \sigma^{\mathrm{DC}} \sim 0.6 \times 10^{-7}\, \frac{\mathrm{W}}{\mathrm{K}} \approx 0.2 \kappa_l\,,
	\label{eqn:kappa}
\end{equation}
where we took $T = 300\,\mathrm{K}$ and
$\sigma^{\mathrm{DC}} = (2 k_F l)(e^2 / h) \sim 200 e^2 / h$,
corresponding to the transport mean-free path of $l \sim 250\, \mathrm{nm}$
at electron density $n = 5 \times 10^{12}\,\mathrm{cm}^2$.
Second, to estimate $g$ and $l_e$, we 
assume that $T$ is higher than the Bloch-Gr\"uneisen temperature
\begin{equation}
	T_{\mathrm{BG}} = 2 \hbar k_F c = 54\,\mathrm{K}\times
	\sqrt{n \,/\, (10^{12}\,\mathrm{cm}^{-2})}.
	\label{eqn:T_BG}
\end{equation}
In this regime the electron cooling power
$g$ due to scattering by acoustic phonons is given by~\cite{Bistritzer2009,Sohier2014}
\begin{equation}
	g = \frac{\pi}{2}\, \frac{n^2 \beta^2}{\hbar \rho v_F^2}
	= \frac{3}{4\pi^2}\,
	\frac{c_v}{\tau_\mathrm{e-ph}}
	\, \frac{T_{\mathrm{BG}}^2}{T^2}\,,
	\label{eqn:g_eph}
\end{equation}
where $c$ is the sound velocity,
$\beta$ is the coupling constant for the strain-induced effective gauge potential, $\rho$ is the graphene mass density per unit area, $c_v$ is the electron specific heat per unit area.
In the second equation in Eq.~\eqref{eqn:g_eph} we
introduced the momentum relaxation time $\tau_\mathrm{e-ph}$
due to electron-phonon scattering,
which enables us to relate the length scales $l_e$ and $l$.
Indeed, from Eqs.~\eqref{eqn:kappa} and \eqref{eqn:g_eph},
we obtain
\begin{equation}
	l_e = 2.6 \sqrt{v_F \tau_\mathrm{e-ph} l}\, \frac{T_{\mathrm{BG}}}{T}\,.
	\label{eqn:l_e_from_tau}
\end{equation}
Assuming that the DC conductivity is dominated by electron-phonon scattering,
so that $l \approx v_F \tau_\mathrm{e-ph}$, we estimate $l_e \approx 1\,\mu\mathrm{m}$ at $T = 300\,\mathrm{K}$.
The length $l_e$ is much larger than $l$ because
at $T \gg T_{\mathrm{BG}}$ the electron-phonon scattering is quasi-elastic.
The net cooling length is determined by the larger of $l_e$ and $l_l$.
In the present case,
\begin{equation}
	l_c \equiv \mathrm{max}(l_e, l_l) = l_e \gg l_l\,.
	\label{eqn:l_c}
\end{equation}
Finally, we find $l_s
= 50\, \mathrm{nm}$ for an SiO$_2$ substrate, which is significantly shorter than $l_l$ and $l_e$.
The electronic cooling length $l_e$ is determined mostly by the doping of the graphene and the electron-phonon coupling, whereas $l_l$ also depends on the interfacial resistance between the graphene and the substrate. 
We expect the effect of the substrate to have the largest impact on the thermal transport properties of the system through the length $l_s$; for a conducting substrate like gold, this length could exceed $l_e$. 
In this work, we consider only the case $l_s\ll l_e$. 
For such interrelations among these characteristic length scales, the behavior of the excess electron temperature $\Delta T = T - T_0$ generated by a local heat source is sketched in Fig.~\ref{Fig:Appendix_schematic}(b). 
In particular, at short distances, $\Delta T$ is described by the equation
\begin{equation}
	\Delta T(\vec{r}) \propto K_0\left(\frac{r}{l_c}\right),
	\label{eqn:T_GF_near}
\end{equation}
where $K_0(z)$ is the MacDonald function.
This temperature profile has a narrow region of exponential decay for $r > l_c$. 
At large distances, however, we find the inverse-distance law
\begin{equation}
	\Delta T(\vec{r}) \propto \frac{1}{r}\,,
	\quad r\gg l_c \ln \frac{l_c}{l_s}.
	\label{eqn:T_GF_far}
\end{equation}
Our derivation of these formulas is presented in Appendix~\ref{appendix:temp_prof}. 

Having defined a model for the temperature, we can now extract the parameter ${\partial\sigma^\mathrm{DC}} / {\partial T}$ in Eq.~\eqref{eqn:I_bolo} experimentally by
measuring non-Ohmic corrections to the current as a function of $V$ in the absence of light. 
Using our model for heat transfer Eq.~\eqref{eqn:T_GF_near}, the corresponding correction is of the form 
\begin{equation}
	\label{eqn:OhmLaw}
	I = \frac{V}{R_g + R} \left(1 - \frac{V^2}{V_0^2} \right)\,,
	\quad V_0^2 = \frac{\kappa}{l_c^2}\, L^2 \left(\frac{\partial\sigma^\mathrm{DC}}{\partial T}\right)^{-1}.
\end{equation}
The cubic nonlinearity predicted by this formula implies that when $V$ is periodically modulated at some small frequency $\Omega$, the current $I$ contains the third harmonic of this frequency.
Measuring this $3\Omega$-signal can then be used to obtain the bolometric coefficient ${\partial\sigma^\mathrm{DC}} / {\partial T}$. 

To generate a net PT photocurrent,
a spatially inhomogeneous thermopower $S$ is required.
We limit ourselves to one-dimensional (1D) inhomogeneities and consider three examples: $S(x)$ having a sharp peak, $S(x)$ exhibiting a step-like change, and $S(x)$ being a linear function of a spatial coordinate $x$.
Such profiles of $S(x)$ can originate from a stacking defect (domain wall) in a multilayered material,
a doping inhomogeneity (e.g, a $p$-$n$ junction in graphene),
or self-gating in a voltage-biased device.
If the dc conductivity $\sigma^\mathrm{DC}(x)$ is a slowly varying function of position, the local thermoelectric coefficient $\sigma^\mathrm{DC} S$ can be estimated from Eq.~\eqref{eqn:Mott}. 
If the source and drain contacts are long conducting strips (Fig.~\ref{fig:heterostructure_schematic}) located at $x = 0$ and $x = L$, the function $\psi_i$ [Eq.~\eqref{eqn:shockley-ramo}] does not have a $y$-component, and
the PT photocurrent can be written in the form
\begin{equation}
	I_\mathrm{PT} 
	 = -\frac{1}{L} \int\limits_0^L \sigma^\mathrm{DC}(x) S(x)\, \partial_x \bar{T}(x) d x\,,
	\label{eqn:I_PC_2D}
\end{equation}
where $\bar{T}(x)$ is the line-integrated excess temperature
\begin{align}
	\bar{T}(x) \equiv \int\! \Delta T(x, y) d y\,.
	\label{eqn:PT_line}
\end{align}
Assuming the tip is far from the sample edges or contacts, we can assume the translationally invariant form $\bar{T}(x, x_t) = \bar{T}(x - x_t)$.

Suppose now that function $S(x)$ has a sharp dip of characteristic depth $\Delta S$ and width $w \ll l_c$ that we can approximate by $S(x) = -w \Delta S\, \delta(x) + S_0$.
This is reasonable for, e.g., domain wall (DW) defect whose width is typically much smaller than the cooling length. 
Substituting this $S(x)$ into the equations above, we obtain
\begin{equation}
	\label{eqn:I_PC_1D_Dlta}
	I^\mathrm{PT}_\mathrm{DW}(x_t)= \sigma^\mathrm{DC}\,\frac{w}{L}\, \Delta S\, \partial_x \bar{T}(-x_t).
\end{equation}
This expression was used to obtain Fig.~\ref{fig:HM_schematic}(b). 
Next, for the thermopower profile
$S(x) = \Delta S\, \Theta(x) + S_0$
characteristic of a $p$-$n$ junction, we find
\begin{equation}
	I^\mathrm{PT}_{p\mathrm{-}n}= \sigma^\mathrm{DC}\, \frac{\Delta S}{L}\, \bar{T}(-x_t).
	\label{eqn:I_PC_1D_Step}
\end{equation}
The plot in Fig.~\ref{fig:Device_Plots}(a) was obtained using the equation above, together with Eqs.~\eqref{eqn:Joule_H}, \eqref{eqn:circular_wave}, and Eq.~\eqref{eqn:T_GF_near}.
Finally, for a linear profile,
$S(x) = (\Delta S / L) x + S_0$, we get
\begin{equation}
	\label{eqn:I_SG_n_edge}
	I^\mathrm{PT}_\mathrm{SG} = \sigma^\mathrm{DC}\, \frac{\Delta S}{L^2}
	\int\limits_{0}^{L} \bar{T}(x - x_t) \, d x\,.
\end{equation}
which has the same $x_t$ dependence as Eq.~\eqref{eqn:I_bolo}.  
In deriving all these results we assumed, for simplicity, that the temperature of the contacts and the adjacent graphene regions is maintained at the ambient value $T_0$,
and so the possible difference in the thermopower of the contacts and graphene does not contribute to $I^\mathrm{PT}$.

\section{Conclusion}
\label{sec:conclusion}
In this paper, we demonstrated applications of several minimal models for scanning near-field photocurrent measurements
on graphene-based heterostructures where effects of plasmon- and phonon-polaritons may be important.
Such collective modes can generate interference patterns near sample edges and other inhomogeneities and exhibit distinctive spectral resonances.
Our models reproduce these interference patterns and elucidate the role of the thermal properties of the heterostructure on the collected signal.
We also studied a coherent photovoltaic contribution to the photocurrent induced by the scanned probe. 
Additionally, we derived a simple relation connecting the frequency dependence of these measurements to that of the parent technique of s-SNOM for the case where dielectric losses in the substrate are negligible.

In this work, we focused mostly on effects that are quadratic in applied electric field. 
On the other hand, a non-perturbative regime of strong applied bias~\cite{Dong2021} may be an interesting direction for future investigations. 
The effects of band structure and especially geometrical phases~\cite{Holder2020,Zeng2021} on photocurrent also warrant study in the novel context of near-field measurements.  
We hope that our modeling of collective mode phenomena in the photocurrent response will be useful in these and other future studies. 

\acknowledgements
	
We thank S. S. Sunku, D. Halbertal, Z. Sun, and Y. Dong for discussions that inspired this work. 


%

\appendix
\section{Hydrodynamic equations}
\label{appendix:HD_equations}

The hydrodynamic equations including terms up to second order in the external field are
\begin{align}
	&\frac{\partial  n}{\partial t} + \nabla \cdot \left(n \vec{u}\right) = 0, 
	\label{eqn:chargecont_2}\\ 
	&\frac{\partial\vec{u}}{\partial t} + \left(\vec{u}\cdot\nabla\right)\vec{u} = -\Gamma_d{\vec{u}}  - \frac{1}{m n}\nabla{P} + \frac{e}{m}\left(\vec{E} + \frac{\vec{u}}{c}\times\vec{B}\right),
	\label{eqn:Drude_modified}\\
	&\frac{\partial n_\varepsilon}{\partial t} + \nabla\cdot \vec{q} 
	= \vec{j}\cdot\vec{E} - \Gamma_E n_\varepsilon,
	\label{eqn:energy_cons} 
\end{align}
Here $\vec{u} = \vec{j} / e n$ is the flow velocity, $P$ is the pressure, and $n_\varepsilon$ is the energy density.
For simplicity, we treat $\Gamma_d$ and the energy relaxation rate $\Gamma_E  \ll \Gamma_d$ as $T$-independent constants, and so our model misses a possible BM effect.
The hydrodynamic mass is $m = \hbar k_F / v_F$, where $v_F$ is the Fermi velocity, $k_F = |4 \pi n / g|^{1 / 2}$ is the Fermi momentum, and $g$ is the spin-valley degeneracy ($g = 4$ in graphene).
We neglect viscosity of the electron fluid in Eq.~\eqref{eqn:Drude_modified} because it affects the results only to the order $O(k^2)$ where $k$ is a characteristic momentum, assumed to be a small quantity.

Using the Gibbs-Duhem relation, the pressure gradient in Eq.~\eqref{eqn:Drude_modified} can be related to the temperature and chemical potential gradients 
\begin{equation}
	\nabla P = s \nabla T + n \nabla \mu\,,
	\label{eqn:P_gibbsduhem}
\end{equation}
where $s$ is the entropy per particle.
The first term contains the Seebeck coefficient and the PTE, Eq.~\eqref{eqn:j_PT}. 
The chemical potential gradient can be split into the $n$- and $T$-dependent parts: 
\begin{equation}
	\nabla\mu = \left(\frac{\partial \mu}{\partial n}\right)_{T} \nabla n+ \left(\frac{\partial \mu}{\partial T}\right)_{n}\nabla T.
	\label{eqn:mu_expansion}
\end{equation}
The first term is responsible for diffusion which gives a contribution proportional to $k^2$, and the second term is the PVT, c.f. Eq.~\eqref{eqn:j_PVT}. 
We also have the equations
\begin{equation}
	d n_\varepsilon = n c_V d T
	\label{eqn:q}
\end{equation}
where $c_V$ is the specific heat per particle and
\begin{equation}
	\vec{q} = -\kappa \nabla T
	+ \left(n_\varepsilon + \frac12 m n u^2\right) \vec{u}\,,
	\label{eqn:n_E}
\end{equation}
where $\kappa$ is the thermal conductivity.

Letting the field be of the form $\vec{E}(\vec{r}, t) = \vec{E}(\vec{r}) e^{-i\omega t} + \mathrm{c.c.}$, the terms linear in $\vec{E}$ give
\begin{align}
	-i \omega\, \vec{j}^{(1)} = -\Gamma_d\, {\vec{j}}^{(1)} + \frac{e^2 n}{m}\, \vec{E}\, ,
	\label{eqn:Drude_def}
\end{align}
which results in the Drude conductivity Eq.~\eqref{eqn:sigma_AC} of the main text. 

We first derive the form of the PVC current from  Eq.~\eqref{eqn:Drude_modified}. 
We assume an incident field of the form $\vec{E}(\vec{r}, t) = \vec{E}_1 e^{-i\vec{k}_1\cdot\vec{r} + i\omega t} + \vec{E}_1 e^{-i\vec{k}_2\cdot\vec{r} - i\omega t} + \mathrm{c.c.}$
Straightforward algebraic manipulations (see  Ref.\cite{Sun2018}) give the dc current as 
\begin{equation}
	\begin{split}
		j_i^{\mathrm{PVC}} &= \frac{e^3 n}{2 m^2(\omega^2 + \Gamma_d^2)}\left[\delta_{lm}\left(\frac{k_{1i}+ k_{2i}}{i\Gamma_d}\right)\right. \\ & \left. + \frac{1}{\omega}\delta_{im}\left(k_{1l} + k_{2l}\right)-\frac{1}{\omega}\delta_{il}\left(k_{1m} + k_{2m}\right) \right. \\ & \left. +\frac{1}{\omega}\delta_{lm}\left(k_{1i} - k_{2i}\right)\right]E_{1l}E_{2m}\, , 
	\end{split}
\end{equation}
which is equivalent to Eq.~\eqref{eqn:j_PVC} of the main text. 

Our result for the current are not yet final because we still need to 
find the time-averaged electron temperature, which can be done using Eqs.~\eqref{eqn:q} and \eqref{eqn:n_E}.  
To this end we need to solve the equation
\begin{equation}
	-\kappa \nabla^2 T + n c_V \Gamma_E  (T - T_0)
	= \left\langle \vec{j}\cdot\vec{E} \right\rangle.
	\label{eqn:temp_hydro}
\end{equation} 
The angular brackets $\left\langle \cdots \right\rangle$ denote the time average.
Note that the time derivative drops out in the DC limit.
Since we neglect terms quadratic in momentum $k$ and higher than second order in field, we can drop the entire $\nabla \cdot \vec{q}$ term as well
and obtain the simple solution
\begin{equation}
	\Delta T \equiv T - T_0 = \frac{2}{\Gamma_E}\, \re \sigma(\omega) E_i^* E_i\,.
	\label{eqn:T_simple}
\end{equation} 
At this point, we can express the second-order current solely in terms of the incident electric field,
that is, we can determine the second-order nonlinear conductivity, which is given by Eq.~\eqref{eqn:sigma^X} of the main text.  


\section{Heat kernel}
\label{appendix:temp_prof}
We consider a system of electrons in graphene thermally coupled to a phonon bath, with the latter in contact with a three-dimensional substrate. 
The differential equations describing this system are 
\begin{align}
	&-\kappa \nabla^2 T(\vec{r}) + g\left[T(\vec{r}) - T_l(\vec{r})\right] = \vec{j}\cdot \vec{E}, \label{eqn:T_elec}\\
		\begin{split} 
		&-\kappa_l \nabla^2 T_l(\vec{r}) + g\left[T_l(\vec{r})- T(\vec{r})\right] \\
		& \qquad + \gamma\left[T_l(\vec{r}) - T_s(\vec{r}, 0)\right] = 0,
		\end{split} \label{eqn:T_phonon}\\
	&-\kappa_s\left(\nabla^2 + \partial_z^2\right)T_s(\vec{r}, z) = 0.\label{eqn:T_sub} 
\end{align}
We will find the Green's function for the electronic temperature $T$ using a two-dimensional Fourier transform. 
The equations \eqref{eqn:T_elec}--\eqref{eqn:T_sub} are 
\begin{align}
	&\kappa q^2 \tilde{T}(\vec{q}) + g\left[\tilde{T}(\vec{q}) - \tilde{T}_l(\vec{q})\right] =  \tilde{P}, \label{eqn:FT_elec}\\
		\begin{split}
	&\kappa_l q^2 \tilde{T}_l(\vec{q}) + g\left[\tilde{T}_l(\vec{q})- \tilde{T}_e(\vec{q})\right]+ \\ 
	 &\qquad \gamma\left[\tilde{T}_l(\vec{r}) - \tilde{T}_s(\vec{q}, 0)\right] = 0, \label{eqn:FT_phonon}
 		\end{split} \\
	\begin{split}
	&\kappa_s\left(q^2 - \partial_z^2\right)\tilde{T}_s(\vec{q}, z) = 0,\\ &{\kappa_s\partial_z \tilde{T}_{s}(\vec{q}, z)\big|_{z = 0} + \gamma \tilde{T}_{s}(\vec{q}, 0) = \gamma \tilde{T}_{l}(\vec{q})}.\label{eqn:FT_sub}
	\end{split}
\end{align}
We first consider the case of constant substrate temperature $T_s(\vec{q}, 0) = T_0$. 
We consider a point source $\tilde{P} = 1$ in Eq.~\eqref{eqn:FT_elec} and change variables $q r = u$. 
Using the expressions for the cooling lengths introduced in Eq.~\eqref{eqn:l_e},we find
\begin{equation}
	\tilde{T}(u) = \frac{1}{\kappa}\frac{u^2 + \frac{r^2}{l_l^2}}{\left(u^2 + u_{-}^2\right)\left(u^2 + u_+^2\right)},  \label{eqn:T_elec_GF} 
\end{equation}
with the roots  
\begin{equation}
	u_{\pm}^2 = \frac{1}{2} \left[\frac{r^2}{l_l^2} + \frac{r^2}{l_e^2} \pm \sqrt{\left(\frac{r^2}{l_e^2} - \frac{r^2}{l_l^2}\right)^2 +  \frac{4gr^4}{l_e^2\kappa_l}}\right]. 
	\label{eqn:T_roots}
\end{equation}
In the case $l_l\gg l_e$ or vice-versa, $l_c = \mathrm{min}(l_e, l_l)$, so 
\begin{equation}
	\tilde{T}(u)\approx \frac{1}{\kappa} \frac{1}{u^2 + \frac{r^2}{l_c^2}}.
	\label{eqn:T_2D} 
\end{equation}
The inverse Fourier transform is
\begin{equation}
	T(\vec{r}) = \frac{1}{2\pi}\int_0^\infty du \,  u \tilde{T}(u) \, J_0(u) = \frac{1}{2\pi\kappa} K_0\left(\frac{r}{l_c}\right). 
	\label{eqn:T_2D_r}
\end{equation}
If $l_e \approx l_l = l_c$, Eq.~\eqref{eqn:T_2D_r} still holds provided that $g\ll\gamma$. 

If one allows for an inhomogeneous temperature in the substrate, Eq.~\eqref{eqn:T_sub} must be included.
We consider the experimentally relevant case $l_l = l_e = l_c\gg l_s$, where $l_s$ is defined in Eq.~\eqref{eqn:l_s}.
Straightforward algebraic manipulations give the following representation of the temperature in the Fourier domain for a point source:
\begin{equation}
	\tilde{T}(u) = \frac{1}{\kappa u}\frac{\left(u^2\frac{l_c^2}{r^2} + \frac{g}{\gamma}\right)\left(1 + u\frac{l_s}{r}\right) + u\frac{l_s}{r}}{\left(1 + u\frac{l_s}{r}\right)\left(u^3\frac{l_s^2}{r^2} + u + \frac{g}{\gamma}\right)+u^2\frac{l_s}{r} + \frac{r l_s}{l_c^2}}. 
	\label{eqn:TP_ratio}
\end{equation}
If $r\ll l_c, l_s$, then Eq.~\eqref{eqn:TP_ratio} is 
\begin{equation}
	\tilde{T}(u) \approx \frac{1}{\kappa}\frac{1}{u^2}, \, T(\vec{r}) = -\frac{1}{2\pi\kappa}\ln r, \, r\ll l_s, l_c. 
	\label{eqn:Tcase1}
\end{equation}
For $r\gg l_s$, we neglect terms containing $\frac{l_s}{r}$ to obtain 
\begin{equation}
	\tilde{T}(u) \approx \frac{1}{\kappa}\frac{u}{u^3 + u\frac{r^2}{l_c^2} + \frac{r^3 l_s}{l_c^4}}, \, l_s\ll r. 
	\label{eqn:Tcase2}
\end{equation}
For $r\ll l_c$, the last term in the denominator is negligible, and we recover Eq.~\eqref{eqn:T_2D_r}. 
As $r$ increases past $l_c$, this term is no longer small.
Taking Eq.~\eqref{eqn:TP_ratio} for $r\gg l_c$, we find 
\begin{equation}
	\tilde{T}(u) \approx \frac{1}{\kappa u}\frac{1}{u + \frac{r \kappa_s}{\kappa}}, \, r\gg l_c. 
\end{equation}
The Fourier transform is 
\begin{equation}
	T(\vec{r}) = \frac{1}{4\kappa}\left[\mathbf{H}_0\left( \frac{r \kappa_s}{\kappa}\right) - Y_0\left(\frac{r \kappa_s}{\kappa}\right)\right], 
\end{equation}
where $\mathbf{H}_\nu(z), Y_\nu(z)$ are the Struve function of the first kind and the Bessel function of the second kind of order $\nu$, respectively.

\end{document}